\documentclass[12pt]{article}
\usepackage{amsmath,amssymb}
\usepackage{bm}

\catcode`\@=11

\ifx\documentclass\undefined
 \def\@sect#1#2#3#4#5#6[#7]#8{\ifnum #2>\c@secnumdepth
     \let\@svsec\@empty\else
     \refstepcounter{#1}\edef\@svsec{\csname prefix#1\endcsname
        \csname the#1\endcsname\hskip 1em}\fi
     \@tempskipa #5\relax
      \ifdim \@tempskipa>\z@
        \begingroup #6\relax
          \@hangfrom{\hskip #3\relax\@svsec}{\interlinepenalty \@M #8\par}%
        \endgroup
       \csname #1mark\endcsname{#7}\addcontentsline
         {toc}{#1}{\ifnum #2>\c@secnumdepth \else
                      \protect\numberline{\csname the#1\endcsname}\fi
                    #7}\else
        \def\@svsechd{#6\hskip #3\relax  
                   \@svsec #8\csname #1mark\endcsname
                      {#7}\addcontentsline
                           {toc}{#1}{\ifnum #2>\c@secnumdepth \else
                             \protect\numberline{\csname the#1\endcsname}\fi
                       #7}}\fi
     \@xsect{#5}}
\else
    \def\@seccntformat#1{\csname prefix#1\endcsname
        \csname the#1\endcsname\quad}
\fi

\@addtoreset{equation}{section}   
\def\theequation{\arabic{section}.\arabic{equation}}

\def\thebibliography#1{\section*{References\@mkboth
 {REFERENCES}{REFERENCES}}\list
 {\leftbibmark\arabic{enumi}\rightbibmark}{
 \settowidth\labelwidth{\leftbibmark #1\rightbibmark}\leftmargin\labelwidth
 \advance\leftmargin\labelsep
 \usecounter{enumi}}
 \def\newblock{\hskip .11em plus .33em minus -.07em}
 \sloppy\clubpenalty4000\widowpenalty4000
 \sfcode`\.=1000\relax}

\def\@citex[#1]#2{\if@filesw\immediate\write\@auxout{\string\citation{#2}}\fi
  \def\@citea{}\@cite{\@for\@citeb:=#2\do
    {\@citea\def\@citea{,\penalty\@m\ }\@ifundefined
       {b@\@citeb}{{\bf ?}\@warning
       {Citation `\@citeb' on page \thepage \space undefined}}%
\hbox{\csname b@\@citeb\endcsname\citemarkdelim}}}{#1}}

\def\@cite#1#2{\leftcitemark{#1 \if@tempswa , #2\fi}\rightcitemark}
\def\leftcitemark{[}
\def\rightcitemark{]}
\def\citemarkdelim{}
\def\leftbibmark{[}
\def\rightbibmark{]}

\catcode`\@=12

\usepackage{amsmath}
\usepackage{amssymb}

\begin{document}

\begin{titlepage}

\vspace{2cm}

\begin{center}\large\bf
Long wavelength limit of 
evolution of cosmological perturbations \\
in the universe where scalar fields and fluids coexist
\end{center}

\begin{center}
Takashi Hamazaki\footnote{email address: yj4t-hmzk@asahi-net.or.jp}
\end{center}

\begin{center}\it
Kamiyugi 3-3-4-606 Hachioji-city\\
Tokyo 192-0373 Japan\\
\end{center}

\begin{center}\bf Abstract\end{center}

We present the LWL formula which represents the long wavelengh limit 
of the solutions of evolution equations of cosmological perturbations 
in terms of the exactly homogeneous solutions 
in the most general case where multiple scalar fields and 
multiple perfect fluids coexist. 
We find the conserved quantity which has origin in the adiabatic 
decaying mode, and by regarding this quantity as the source term
we determine the correction term which corrects the discrepancy between 
the exactly homogeneous perturbations and the $k \to 0$ limit of 
the evolutions of cosmological perturbations.
This LWL formula is useful for investigating the evolutions
of cosmological perturbations in the early stage of our universe 
such as reheating after inflation and the curvaton decay 
in the curvaton scenario.
When we extract the long wavelength limits of evolutions of 
cosmological perturbations from the exactly homogeneos perturbations 
by the LWL formula,
it is more convenient to describe the corresponding exactly homogeneous system
with not the cosmological time but the scale factor as the evolution
parameter.
By applying the LWL formula to the reheating model and the curvaton
model with multiple scalar fields and multiple radiation fluids,
we obtain the S formula representing the final amplitude of the 
Bardeen parameter in terms of the initial adiabatic and isocurvature
perturbations

PACS number(s):98.80.Cq

\end{titlepage}

\section{Introduction and summary}

Recently we come to be required to investigate the evolution of
cosmological perturbations in the very early universe 
\cite{Polarski1992}, \cite{Gordon2001}.
According to the inflationary scenarios and the curvaton scenario,
in the early universe
the wavelength of cosmological perturbations responsible for
the present cosmic structures such as galaxies and clusters of galaxies
is much larger than the horizon scales.
Therefore the methods for researching the cosmological perturbations
on superhorizon scales have been sought.
In this context, Nambu and Taruya pointed out that there exists an LWL
formula representing the $k \to 0$ limit of the cosmological
perturbations in terms of the exactly homogeneous perturbations
\cite{Taruya1998}.
Soon later in the multiple scalar fields system \cite{Kodama1998}
\cite{Sasaki1998}, the complete LWL formulae were constructed.
Since the evolution equations of the corresponding exactly homogeneous
universe look simpler than the evolution equations of cosmological
perturbations, the LWL formula brought about great simplification.
In addition, the viewpoint that the evolutions of the cosmological 
perturbations on superhorizon scales are governed by the stability
and instability of the corresponding exactly homogeneous universe 
\cite{Hamazaki2002}, \cite{Hamazaki2004}
is useful for physical interpretation.
In this context, the phase space of the corresponding exactly
homogeneous system was investigated in detail and the role
of the fixed points in the phase space in the stability and 
instability of cosmological perturbations was discussed 
\cite{Hamazaki2004}.
For these reasons, the LWL formula was used for investigating the
evolution of cosmological perturbation on superhorizon scales by
several authors \cite{Kodama1996}, \cite{Sasaki1998}, \cite{Taruya1998},
\cite{Hamazaki2002}, \cite{Hamazaki2004}.
In this paper, in section $2$, in order to investigate the evolutionary
behaviors of cosmological perturbations during reheating and the curvaton
decays, we construct the complete LWL formulae 
for the most general system where the multiple scalar fields and the 
multiple perfect fluids coexist.

In the early universe, 
the cosmological perturbations on superhorizon scales responsible for 
the cosmic structures and the CMB temperature anisotropies experience
the reheating and/or the curvaton decays. 
In these processes, the multiple scalar fields such as inflatons and 
curvatons oscillate coherently, gradually decaying into radiation fluid.  
By replacing the oscillating scalar fields with the dust fluids,
the evolution of the cosmological perturbations during 
reheating after the inflation \cite{Hamazaki1996} and 
in the curvaton decay \cite{Malik2003} were investigated.
These authors treated the system dominated by dust-like scalar field
fluid and radiation and investigated the influence of the entropy
perturbation originating from the multicomponent property to the
evolution of the total curvature perturbation variables such as 
the Bardeen parameter.
The purpose of these analyses was to determine the initial perturbation
of the present Friedmann universe 
in terms of the early stage seed perturbation.
Although it was shown partially that this replacement is physically
reasonable \cite{Kodama1996}, \cite{Hamazaki1996}, we are required to
treat the decaying oscillatory scalar fields directly.
In fact, the instabilities characteristic to 
the rapidly oscillating scalar fields were pointed out 
\cite{Bassett2000}, \cite{Finelli2000}, \cite{Zibin2001}, 
\cite{Hamazaki2004}, \cite{Yoshida2004} and  investigated
\cite{Traschen1990}, \cite{Kofman1994}, \cite{Shtanov1995}, 
\cite{Kofman1997}.
In order to treat the oscillatory scalar fields directly,
the action angle variables were introduced and the averaging method
representing the averaging over the fast changing angle variables
was applied \cite{Hamazaki2002},\cite{Hamazaki2004}.
In this paper, in order to investigate the evolutionary behaviors of
cosmological perturbations during reheating and/or the curvaton decays,
in section $3$ the action angle variables and the action angle
perturbation variables are introduced, and in section $5$, the averaging 
method was applied to the decaying oscillatory scalar fields.

In the papers \cite{Hamazaki2002}, \cite{Hamazaki2004}, the averaging
method was used to investigate the corresponding exactly homogeneous system,
and by the LWL formula the evolution of the cosmological perturbation
in the long wavelength limit was constructed from the corresponding
exactly homogeneous perturbation.
By using the LWL formula and by applying the averaging method,
in single oscillatory scalar field system \cite{Kodama1996} and 
in nonresonant multiple oscillatory scalar fields system \cite{Hamazaki2002}  
it was shown that the Bardeen parameter is conserved, 
and in resonant multiple oscillatory scalar fields system \cite{Hamazaki2004}
it was shown that the cosmological perturbation including the Bardeen 
parameter can grow.
In this paper, in section $6$, $7$, 
by using the LWL formula and by applying the averaging method to the
decaying oscillatory scalar fields, we construct S formulae representing
the final amplitude of the Bardeen parameter in terms of the initial
seed adiabatic and entropic perturbations, in the reheating and in the
curvaton decays, respectively.

The organization and the summary of the paper is explained as follows.
In section $2$, we construct the LWL formula as for 
such scalar-fluid composite system based on the the philosophy
of the paper \cite{Kodama1998}.
The discrepancy exists between 
the evolution equations of the cosmological perturbations in 
the $k \to 0$ limit and the evolution equations of the exactly 
homogeneous perturbations because the former contains $k^2 \Phi = O(1)$
terms and the latter does not, therefore this discrepancy should be
corrected by the correction term which contributes 
the adiabatic decaying mode,
but any general methods for determining such correction term have not 
been presented yet, and only in the multiple scalar fields system 
such correction term was determined.
We show in the $k \to 0$ limit the existence of the conserved quantity
which has origin in the adiabatic decaying mode and which is related
with $k^2 \Phi$.
By regarding this conserved quantity as the source term, and obtaining the 
special solution $A^{\flat}$, we correct the exactly homogeneous
perturbation $A^{\sharp}$ and we obtain the complete LWL formula
$A = A^{\sharp} + A^{\flat}$ in the most general scalar-fluid composite
system.
In section $3$, we point out that it is more appropriate to use 
the scale factor $a$ rather than the cosmological time $t$ as the 
evolution parameter when we use the LWL formula.
As for the scalar quantity $T$, we use the perturbation variable $DT$,
$D$ is the operator which maps the exactly homogeneous scalar quantity
$T$ to the gauge invariant perturbation variable representing the $T$
fluctuation in the flat slice.
$D$ defined in this way can be interpreted as a kind of derivative
operator.
In fact, the exactly homogeneous part $(D T)^{\sharp}$ can be expressed
as the derivative of $T$ with respect to the solution constant with the 
scale factor $a$ fixed.
In order to investigate the exactly homogeneous system containing 
oscillatory scalar fields, we use the action angle variables $I_a$,  
$\theta_a$.
By using $D$ defined in this way, we can define the action angle 
perturbation variables $D I_a$, $D \theta_a$ whose exactly homogeneous 
parts are given as the derivatives of $I_a$, $\theta_a$ with solution 
constant $C$ with the scale factor $a$ fixed.
When we use the derivative operator $D$ and the LWL formulae,
it is essential to use the scale factor $a$ as the evolution parameter.
In section $4$, we apply the LWL formulae to the non-interacting
multicomponents system and discuss the long wavelength limit of
the evolution of the Bardeen parameter.
In section $5$, we apply the averaging method by which the system
is averaged over the fast changing angle variables to the decaying 
scalar fields which have been discussed in the reheating model 
and the curvaton model.
By evaluating the corrections produced by the averaging process
and the errors produced by the truncation of the sufficient reduced
angle variables dependent part, the validity of the averaging method
is established.
In section $6$, $7$, we apply the LWL formula and the averaging method
to the interacting multicomponents model such as the reheating model,
the curvaton model, respectively.
We assume that the multiple scalar fields and the multiple radiation 
fluid components exist.
In these models, we construct the S formulae representing the 
final amplitude of the Bardeen parameter in terms of the initial
adiabatic and isocurvature perturbations.
In our previous paper \cite{Hamazaki2004}, the evolutionary behaviors
of cosmological perturbations in the early universe where multiple
oscillatory scalar fields interact with each other have been 
investigated.
This S formula give the information about how the cosmological 
perturbations which grew in such early era are transmitted into
the radiation energy density perturbations through the 
energy transfer from the scalar fields into the radiation fluids.
We present the necessary condition for the initial entropic
perturbations produced in the early era to survive until the 
late radiation dominant universe.
Section $8$ is devoted to discussions containing non-linear
generalization of our LWL formailism and comment of
the case where the decay rate depends on other physical quantities.
In appendices, the proofs of the propositions presented in section $5$
and the evaluations of the useful mathematical formulae used in 
section $6$ are contained.

In this paper, we consider the case where the homogeneous 
scalar fields obey the phenomenological evolution equations as
\begin{equation}
 \ddot{\phi} + 3 H \dot{\phi} + \frac{\partial U}{\partial \phi} + S = 0.
\end{equation}
The interactions between scalar fields are described by the interaction
potential $U$, while the interaction between scalar fields and fluids
are described by $S$.
This analysis includes the well known case $S = \Gamma \dot{\phi}$ 
\cite{Hosoya1984}, \cite{Morikawa1986} and
the general case where $S$ is an arbitrary analytic function of
$\phi$, $\dot{\phi}$ which was discussed in the paper
\cite{Yokoyama2004} but whose perturbations have not been investigated
yet.
Another supplemental purpose of our paper is to present the evolution
equations of cosmological perturbations corresponding to the homogeneous
system with various source term $S$ especially dependent on $\dot{\phi}$.

The notation used in this paper are based 
on the the review \cite{Kodama1984} and 
the paper \cite{Kodama1998}.

\section{Derivation of the LWL formula}

We give the definitions and the evolution equations as for the 
background and the perturbation variables.
Based on these notations, in the most general model where the 
multiple scalar fields and the multiple perfect fluid components
interact, we give the LWL formula representing the evolutions 
of the perturbations variables in terms of the exactly homogeneous
solutions.

We consider perturbations on a spatially flat Robertson-Walker universe
given by
\begin{eqnarray}
 \tilde{d s}^2 &=& - (1+ 2 A Y) dt^2 - 2 a B Y_j dt dx^j
\notag \\ 
&& + a^2 [ (1 + 2 H_L Y) \delta_{jk} + 2 H_T Y_{jk} ]
  dx^j dx^k,
\end{eqnarray}
where $Y$, $Y_j$ and $Y_{jk}$ are harmonic scalar, vector and tensor
for a scalar perturbation with wave vector $\bm{k}$ on flat three-space:
\begin{equation}
 Y := e^{i \bm{k} \cdot \bm{x}}, \quad 
 Y_j := - i \frac{k_j}{k} Y, \quad
 Y_{jk} := \left( \frac{1}{3} \delta_{jk} - \frac{k_j k_k}{k^2} \right) Y.
\end{equation}
By using the gauge dependent variables ${\cal R}$ and $\sigma_g$
representing the spatial curvature perturbation and the shear,
respectively:
\begin{equation}
 {\cal R} := H_L + \frac{1}{3} H_T, \quad
 \sigma_g := \frac{a}{k} \dot{H}_T -B,
\end{equation}
we can define two independent gauge invariant variables:
\begin{equation}
 {\cal A} := A - \left( \frac{{\cal R}}{H} \right)^{\cdot}, \quad
 \Phi := {\cal R} - \frac{a H}{k} \sigma_g.
\end{equation}
In order to define the matter perturbation variables, we will 
consider the scalar quantity perturbation variables generally.
As for covariant scalar quantity $\tilde{T} = T + \delta T Y$,
we define the gauge invariant perturbation variable representing 
the $T$ fluctuation in the flat slice:
\begin{equation}
 D T := \delta T - \frac{\dot{T}}{H} {\cal R},
\label{dtflatslice}
\end{equation}
Next we consider the covariant scalar quantity $\tilde{T}_2$
whose background quantity is the time derivative of $T$: $\dot{T}$.
The extension of $\dot{T}$ into the covariant scalar quantity
$\tilde{T}_2$ is not unique.
For example,
\begin{equation}
 \tilde{T}_2 = {\rm sgn} \left( \partial_0 \tilde{T} \right)
\left[ - \tilde{g}^{\mu \nu}
 \tilde{\nabla}_\mu \tilde{T} \tilde{\nabla}_\nu \tilde{T}
\right]^{1/2},
\end{equation}
and
\begin{equation}
 \tilde{T}_2 = \tilde{n}^{\mu} \tilde{\nabla}_{\mu} \tilde{T},
\end{equation}
where $\tilde{n}^{\mu}$ is an arbitrary vector field satisfying
\begin{equation}
 \tilde{n}^{\mu} \tilde{n}_{\mu} = -1,
\end{equation}
have the same $\dot{T}$ as the background part.
But these defferent $\tilde{T}_2$'s give the unique perturbation
part: $D T_2 = (D T)^{\cdot} -\dot{T} {\cal A}$.
Therefore we can define $D \dot{T}$ by
\begin{equation}
 D \dot{T} := 
 (D T)^{\cdot} -\dot{T} {\cal A}.
\label{ddotT}
\end{equation}

We consider the the universe where the scalar fields $\phi_a$, 
$(1 \le a \le N_S)$ and the fluids $\rho_\alpha, P_\alpha$ 
$(1 \le \alpha \le N_f)$ coexist, whose energy momentum tensor is 
divided into $A=(S,f)$ parts where $S$ represents the multiple scalar fields,
$f$ represents the multiple fluids.
The energy momentum tensor of $f$ part are further divided into
individual fluids parts $\alpha$.  
On the other hand, the energy momentum tensor of $S$ part cannot be
divided into individual scalar fields parts $a$, 
since the interaction potential $U$
contains the terms consisting of plural scalar fields $\phi_a$:
\begin{eqnarray}
\tilde{T}^\mu_{\> \nu} &=& 
\left(  \tilde{T}^\mu_{\> \nu}\right)_S +
\left(  \tilde{T}^\mu_{\> \nu}\right)_f
= \left(  \tilde{T}^\mu_{\> \nu}\right)_S
+\sum_\alpha \tilde{T}^\mu_{\alpha \nu},
\label{emdivision}\\ 
0 &=&
\left( \tilde{Q}_\mu \right)_S +
\left( \tilde{Q}_\mu \right)_f
= \left( \tilde{Q}_\mu \right)_S
+\sum_{\alpha} \tilde{Q}_{\alpha \mu}, 
\label{transferdivision}
\end{eqnarray}
where the energy-momentum transfer vector $\tilde{Q}_{A \mu}$
is defined by
\begin{equation}
 \tilde{\nabla}_{\nu} \tilde{T}^{\nu}_{A \mu} = \tilde{Q}_{A \mu}
 = \tilde{Q}_{A} \tilde{u}_{\mu} + \tilde{f}_{A \mu},
\end{equation}
where $\tilde{u}_{\mu}$ is the four velocity of the whole matter
system and  the momentum transfer $\tilde{f}_{A \mu}$ satisfies
$\tilde{u}^{\mu} \tilde{f}_{A \mu} = 0$.
For the scalar perturbation, the energy-momentum tensor and 
the energy-momentum transfer vector of each individual component
are expressed as
\begin{eqnarray}
 \tilde{T}^0_{A0} &=& - (\rho_A + \delta \rho_A Y),\\ 
 \tilde{T}^0_{Aj} &=& a (\rho_A + P_A)(v_A - B) Y_j,\\
 \tilde{T}^j_{Ak} &=& 
 (P_A \delta^j_k + \delta P_A Y \delta^j_k + \Pi_{TA} Y^j_k),
\end{eqnarray}
and
\begin{eqnarray}
 \tilde{Q}_{A0} &=& - [Q_A + (Q_A A + \delta Q_A) Y],\\ 
 \tilde{Q}_{Aj} &=& a [Q_A(v - B) +F_{cA}] Y_j,
\end{eqnarray}
where $\rho_A$, $P_A$ and $Q_A$ are the background quantities of 
the energy density, the pressure and the energy transfer of the 
individual component $A$, respectively.
The anisotropic pressure perturbation $\Pi_{TA}$ and 
the momentum transfer perturbation $F_{cA}$ are already gauge invariant.
As for the scalar quantities $T=(\rho_A, P_A, Q_A)$, we use $DT$ as the 
gauge invariant perturbation variables.
As for the gauge invariant velocity perturbation variable, we use
\begin{equation}
 Z_A := {\cal R} - \frac{a H}{k} \left( v_A - B \right).
\end{equation}

The energy-momentum tensor of scalar fields part is given by
\begin{equation}
\left( \tilde{T}^\mu_{\> \nu} \right)_S =
\tilde{\nabla}^\mu \tilde{\phi} \cdot \tilde{\nabla}_\nu \tilde{\phi}
-\frac{1}{2}
\left( \tilde{\nabla}^\lambda \tilde{\phi} \cdot \tilde{\nabla}_\lambda \tilde{\phi} 
+ 2 \tilde{U} \right)
\delta^\mu_\nu.
\end{equation}
Since divergence of the energy momentum tensor is given by
\begin{equation}
\left( \tilde{\nabla}_\mu
\tilde{T}^\mu_{\> \nu} \right )_S =
\left( \tilde{\Box} \tilde{\phi}_a - 
 \frac{\partial \tilde{U}}{\partial \tilde{\phi}_a} \right)
 \tilde{\nabla}_\nu \tilde{\phi}_a,
\end{equation}
in order that the phenomenological equations of 
motion of the scalar fields become
\begin{equation}
 \tilde{\Box} \tilde{\phi}_a - 
 \frac{\partial \tilde{U}}{\partial \tilde{\phi}_a} =
 \tilde{S}_a,
\label{scalarmotion}
\end{equation}
we assume that
\begin{equation}
 \left( \tilde{Q}_\nu \right)_S =
 \tilde{S}_a
 \tilde{\nabla}_\nu \tilde{\phi}_a,
\end{equation}
By using the scalar fields background variables 
$\phi_a$, $\dot{\phi}_a$, $S_a$ 
and the corresponding perturbation variables 
$D \phi_a$, $D \dot{\phi}_a$, $D S_a$, 
the background part of the fluid variables are given by
\begin{eqnarray}
 \rho_S &=& \frac{1}{2} \left( \dot{\phi} \right)^2 + U,\\
 P_S &=& \frac{1}{2} \left( \dot{\phi} \right)^2 - U,\\
 h_S &=& \left( \dot{\phi} \right)^2,\\
 Q_S &=& - S \cdot \dot{\phi},
\end{eqnarray}
and the perturbation part of fluid variables are given by  
\begin{eqnarray}
 \left( D \rho \right)_S
 &=& \frac{\partial \rho_S}{\partial \phi} \cdot D \phi
+ \frac{\partial \rho_S}{\partial \dot{\phi}} \cdot D \dot{\phi},\\
 \left( D P \right)_S
 &=& \frac{\partial P_S}{\partial \phi} \cdot D \phi
+ \frac{\partial P_S}{\partial \dot{\phi}} \cdot D \dot{\phi},\\
 \left( h Z \right)_S 
 &=& - H \dot{\phi} \cdot D \phi,\\
 \left( \Pi_T \right)_S
 &=& 0,\\ 
 \left( D Q \right)_S
 &=& - S \cdot D \dot{\phi} 
     - \dot{\phi} \cdot DS,\label{transferperturb}\\
 \left( a F_c \right)_S
 &=&
 S_a \left( - k D \phi_a - \frac{k}{H} \dot{\phi}_a Z \right).
\end{eqnarray}
When the source of the scalar field $\phi_a$, $S_a$ is given as 
functions of the covariant scalar quantities $\tilde{T}$ and
$\tilde{T}_2$ whose background part is $\dot{T}$, that is
$\tilde{S}_a = \tilde{S}_a (\tilde{T}, \tilde{T}_2)$,
$D S_a$ is given by
\begin{equation}
 DS_a =  \frac{\partial S_a}{\partial T} \cdot D T
 + \frac{\partial S_a}{\partial \dot{T}} \cdot D \dot{T}.
\label{sperturb}
\end{equation}
In such case, $(D Q)_S$ can be written as 
\begin{equation}
 (D Q)_S =  \frac{\partial Q_S}{\partial T} \cdot D T
 + \frac{\partial Q_S}{\partial \dot{T}} \cdot D \dot{T},
\label{qperturb}
\end{equation}
which is assumed from now on.
In the same way as the individual components 
$\tilde{T}^{\mu}_{A \nu}$, as for the total energy-momentum tensor
$\tilde{T}^{\mu}_{\> \nu} = \sum_A \tilde{T}^{\mu}_{A \nu}$,
we can define the gauge invariant perturbation variables such as
$D \rho$, $D P$, $h Z$ and $\Pi_T$.
From (\ref{emdivision}),(\ref{transferdivision}),
we obtain the background equations as
\begin{eqnarray}
 \rho &=& \rho_S + \sum_\alpha \rho_\alpha,\\
 P &=& P_S + \sum_\alpha P_\alpha,\\
 h &=& h_S + \sum_\alpha h_\alpha,\\
 0 &=& Q_S + \sum_\alpha Q_\alpha,
\end{eqnarray}
and perturbation equations as
\begin{eqnarray}
 D \rho &=& D \rho_S
 + \sum_\alpha D \rho_{\alpha},\\
 D P &=& D P_S
 + \sum_\alpha D P_{\alpha},\\
 h Z &=& \left( h Z \right)_S
 + \sum_\alpha h_\alpha Z_{\alpha},\\
 \Pi_T &=& \left( \Pi_{T} \right)_S
 + \sum_\alpha \Pi_{T \alpha},\\
 0 &=& \left( D Q \right)_S
 + \sum_\alpha D Q_\alpha,\\
 0 &=& \left( F_c \right)_S
 + \sum_\alpha F_{c \alpha}.
\end{eqnarray}
This $Z$ is known as the Bardeen parameter \cite{Bardeen1980} 
\cite{Kodama1984}, \cite{Mukhanov1992}.
In the long wavelength limit, the Bardeen parameter is conserved 
in the case where the entropy perturbations are negligible.
But in various systems it was reported that the entropy perturbations
cannot be neglected \cite{Bassett2000}, \cite{Finelli2000},
\cite{Gordon2001}, \cite{Hamazaki2004}, so in the present paper
we will investigate the evolutionary behavior of the Bardeen parameter 
more carefully. 
Until now, as for the gauge invariant scalar quantity perturbation 
variables, we use $D$.
But traditionaly most scalar quantity perturbation variables have been 
written without using $D$:
\begin{equation}
 Y_a := D \phi_a, \quad 
 \rho_\alpha \Delta_{g \alpha} := D \rho_\alpha, \quad
 P_\alpha \Pi_{L \alpha} := D P_\alpha, \quad 
 Q_\alpha E_{g \alpha}:= D Q_\alpha.
\end{equation}
This $Y_a$ has been called the Sasaki-Mukhanov variable
\cite{Sasaki1986} \cite{Mukhanov1988}.

In terms of the gauge independent variables defined above,
we give the evolution equations of cosmological perturbations.
From (\ref{scalarmotion}), the background and the perturbation parts
can be written as
\begin{eqnarray}
 &&
 \ddot{\phi} + 3 H \dot{\phi} + \frac{\partial U}{\partial \phi} + S = 0,
\label{scalarback}\\ 
 && L_1 (DT, {\cal A}) = - \frac{k^2}{a^2} D \phi 
 - \frac{k^2}{a^2} \frac{\dot{\phi}}{H} \Phi,
\label{scalarperturb}
\end{eqnarray}
where
\begin{equation}
 L_1 (DT, {\cal A}) =
 (D \phi)^{\cdot \cdot} + 3 H (D \phi)^{\cdot} 
 + \frac{\partial^2 U}{\partial \phi \partial \phi} D \phi + DS 
 - \dot{\phi} \dot{{\cal A}}
 + 2 (\frac{\partial U}{\partial \phi} +S) {\cal A}.
\end{equation}
As for the fluid components,
$\tilde{\nabla}_\mu \tilde{T}^\mu_{\alpha \nu} = \tilde{Q}_{\alpha \nu}$
gives the background equations as
\begin{equation}
 \dot{\rho}_\alpha = - 3 H h_\alpha + Q_\alpha,
\label{componentback}
\end{equation}
and the perturbation equations as
\begin{eqnarray}
&& L_{2 \alpha} (DT, {\cal A})
= - \frac{k^2}{a^2 H} h_\alpha \left( \Phi - Z_{\alpha} \right),
\label{componentperturb}\\
&& \left( \frac{h_{\alpha} Z_{\alpha} }{H} \right)^{\cdot}
 + 3 h_\alpha Z_\alpha + h_\alpha {\cal A} 
 + D P_{\alpha} - \frac{2}{3} \Pi_{T \alpha}
 = - \frac{a}{k} F_{c \alpha} + \frac{Q_\alpha}{H} Z,
\label{bardeenperturb}
\end{eqnarray}
where
\begin{equation}
 L_{2 \alpha} (DT, {\cal A})
= \left( D \rho_{\alpha} \right)^{\cdot}
 + 3 H D \rho_\alpha
 + 3 H D P_{\alpha} - Q_\alpha {\cal A}
 - D Q_{\alpha}.
\end{equation}
$\tilde{G}^\mu_{\> \nu} = \kappa^2 \tilde{T}^\mu_{\> \nu}$
gives the background equations as
\begin{eqnarray}
 H^2 &=& \frac{\kappa^2}{3} \rho,
\label{hubble}\\
 \dot{\rho} &=& - 3 H h,
\\
 \dot{H} &=& - \frac{3}{2} (1+w) H^2,
\label{hubbledot}
\end{eqnarray}
and the perturbation equations as
\begin{eqnarray}
&& L_3 (DT, {\cal A})
= 2 \rho \frac{k^2}{3 a^2 H^2} \Phi,
\label{poissonperturb}\\
&& L_4 (DT, {\cal A})
= - \frac{\kappa^2}{3} \Pi_T
- \frac{k^2}{a^2} \Phi,
\label{hubbledotperturb}\\
&& {\cal A} + \frac{3}{2} (1+ w) Z =0,
\label{constraint}\\ 
&& {\cal A} + \frac{1}{a} 
\left( \frac{a}{H} \Phi \right)^{\cdot}
= - \frac{\kappa^2}{k^2} a^2 \Pi_T,
\label{newtonpot}
\end{eqnarray}
where 
\begin{eqnarray}
  L_3 (DT, {\cal A})
&=&
 2 \rho {\cal A} + D \rho,\\
 L_4 (DT, {\cal A})
&=&
 H \dot{{\cal A}} + 2 \dot{H} {\cal A}
 - \frac{\kappa^2}{2} 
 \left( D \rho + D P \right).
\end{eqnarray}
The dynamical perturbation variables are classified into 
two groups, that is, what has analogy with the exactly homogeneous 
perturbations and what is not related with the exactlty homogeneous
perturbations at all.
The dynamical perturbation variables of the former type are
$D T$ representing the scalar quantity $T=(\rho, P, \phi, Q, S)$
perturbation in the flat slice, $D \dot{T}$ and 
the metric perturbation variable ${\cal A}$. 
The dynamical perturbation variables of the latter type are the
Newtonian gravitational potential $\Phi$ and
$Z_A$, $F_{c A}$, $\Pi_{T A}$ which have vector or tensor
origin.
In the above $L_i (i=1, \cdot \cdot \cdot, 4)$ equations,
the former type dynamical perturbation variables are contained 
in the left hand side while the latter type perturbation variables
are collected in the right hand side.
The exactly homogeneous perturbations $D T^{\sharp}$ and 
${\cal A}^{\sharp}$ corresponding to $D T$ and ${\cal A}$, respectively
are constructed as
\begin{eqnarray}
 \left( D T \right)^{\sharp} &:=& 
 \left( \frac{\partial T}{\partial C} \right)_t 
- \frac{ \dot{T} }{ H }
 {\cal R}^{\sharp},
\label{dtsharp1}\\
 {\cal A}^{\sharp} &:=&
 - \left( \frac{{\cal R}^{\sharp}}{H} \right)^{\cdot},\\
 {\cal R}^{\sharp} &:=& 
 \frac{1}{a}
 \left( \frac{\partial a}{\partial C} \right)_t,
\label{dtsharp2}
\end{eqnarray}
where $C$ is the solution constant of the background solution and
the subscript $t$ implies that the derivative with respect to $C$
is performed with the cosmological time $t$ fixed.
On the other hand, the dynamical perturbation variables of 
the latter type such as $\Phi$, $Z_A$, $F_{c A}$ and $\Pi_{T A}$ 
do not have exactly homogeneous counterparts.
The evolution equations of cosmological perturbations containing
$L_i (i=1, \cdot \cdot \cdot, 4)$ have analogy in the exactly
homogeneous perturbation equations. 
In fact, the variations of the exactly homogeneous equations  
(\ref{scalarback}), (\ref{componentback}), (\ref{hubble})  
and (\ref{hubbledot}) give
\begin{equation}
 L_i ({DT}^{\sharp}, {\cal A}^{\sharp})=0 \quad
(i=1, \cdot \cdot \cdot ,4),
\label{lvar}
\end{equation}
respectively.
The only difference between the exactly homogeneous perturbation 
$L_i (i=1, \cdot \cdot \cdot, 4)$ equations and the actual $k \neq 0$
cosmological perturbation $L_i (i=1, \cdot \cdot \cdot, 4)$ equations
is that $k^2 \Phi$ terms exist in the latter but $k^2 \Phi$ terms 
do not exist in the former.
Then the effect of the source term $k^2 \Phi$ is corrected in the
following way.
In performimg the correction process, it is important to notice 
that the source terms $k^2 \Phi$ can be represented 
in terms of conserved quantity 
which has origin in the universal adiabatic decaying mode. 
In fact, as for $f$ defined by
\begin{equation}
 f = a^3 H \left( {\cal A} + \frac{1}{2} \Delta_g \right)
   = \frac{k^2}{3 H} a \Phi,
\end{equation}
using (\ref{componentperturb}), (\ref{poissonperturb}),
(\ref{hubbledotperturb}) yields
\begin{equation}
 \frac{d f}{d t} = - a^3 H^2 w \Pi_T + \frac{1}{2} a k^2 (1+w) Z.
\end{equation}
When we assume that for $k \to 0$ limit
\begin{equation}
 \Pi_T \to 0, \quad \quad k Z \to 0,
\label{k0limitcond}
\end{equation}
are satisfied, the quantity $f$ is conserved, whose value is written
as $c$.
Therefore for $k \to 0$ limit,
\begin{equation}
 k^2 \Phi \to \frac{3 H}{a} c = O(1).
\label{conservquantity}
\end{equation}
This expression of $\Phi$ is well known as that of the universal 
adiabatic decaying mode \cite{Kodama1998}.
In the $L_i (i=1, \cdot \cdot \cdot, 4)$ equations containing
$DT$, ${\cal A}$, the Newtonian potential $\Phi$ appears only
in the form $k^2 \Phi$, that is, accompanied by $k^2$.
When we assume that $DT =O(1)$, ${\cal A}=O(1)$, $k^2 \Phi$ behaves
as $O(1)$.
Since in the linear perturbation, the scale of the perturbation
variables is arbitrary, the fact that $\Phi = O(1 / k^2)$ does not
imply the breakdown of the linear perturbation.
If one want to get $\Phi=O(1)$, one simply assumes that 
$DT =O(k^2)$, ${\cal A}=O(k^2)$.
But as explained later, we cannot assume that $\Phi$ is vanishing,
since $c$ defined by (\ref{conservquantity}) must satisfy 
the constraint (\ref{constrainthomosol}).
Therefore in the $k \to 0$ limit where (\ref{k0limitcond}), 
(\ref{conservquantity}) are satisfied,
(\ref{scalarperturb}), (\ref{componentperturb}),
(\ref{poissonperturb}), (\ref{hubbledotperturb})
can be written as
\begin{eqnarray}
  L_{1 a} (DT, {\cal A}) &=&
- \frac{3 \dot{\phi}_a}{a^3} c,
\label{l1c}\\
 L_{2 \alpha} (DT, {\cal A})
&=& - \frac{3 h_\alpha}{a^3} c,
\label{l2c}\\
 L_3 (DT, {\cal A})
&=& \frac{2 \rho}{a^3 H} c,
\label{l3c}\\
 L_4 (DT, {\cal A})
&=& - \frac{3 H}{a^3} c.
\label{l4c}
\end{eqnarray}
It can be verified that above four sets of equations (\ref{l1c}),
(\ref{l2c}), (\ref{l3c}) and (\ref{l4c}) are satisfied by
\begin{eqnarray}
 {\cal A} &=& \frac{3}{2} (1+w) g + \frac{\dot{g}}{H},\\
 DT &=& \frac{\dot{T}}{H} g,
\end{eqnarray}
where
\begin{equation}
 g = c \int_{t_0} dt \frac{1}{a^3}.
\label{gfactor}
\end{equation}
This special solution for
$A=\left(DT, {\cal A} \right)$
is written as $A^{\flat}$.
Since the variation of the exactly homogeneous solution $A^{\sharp}$
satisfies (\ref{lvar}),
the general solutions of (\ref{l1c}), (\ref{l2c}), (\ref{l3c}), 
(\ref{l4c})
$A=\left(DT, {\cal A} \right)$ can be expressed as
\begin{equation}
 A = A^{\sharp} + A^{\flat}.
\end{equation}

The perturbation equations except $L_i (i=1, \cdot \cdot \cdot, 4)$
equations have vector origin, that is, they are derived from the 
space component of the Einstein equations.
Therefore these perturbation equations do not have any analogy
with the exactly homogeneous perturbation equations. 
As explained in the paper \cite{Kodama1998},
these perturbation equations determine the evolutions of the 
dynamical perturbation variables which have vector or tensor origin,
that is, which have no correspondence with the exactly homogeneous 
pertubations, or give the constraint which should be satisfied 
in order that the exactly homogeneous perturbations become the 
$k \to 0$ limit of evolutions of cosmological perturbations. 
Therefore
(\ref{bardeenperturb}), (\ref{constraint}) can be interpreted as 
the decision of the evolution of the variables $Z_\alpha$ which 
is not related to the exactly homogeneous solution at all 
in terms of 
$A=\left(DT, {\cal A} \right)$, the constraint to the exactly 
homogeneous perturbations, respectively.
Integrating (\ref{bardeenperturb}) yields
\begin{equation}
h_{\alpha} Z_{\alpha} \to
\frac{H}{a^3} \left[
 C_\alpha + \int_{t_0} dt a^3 \left(
 - h_\alpha {\cal A} 
 - DP_{\alpha}  
 - \frac{a}{k} F_{c \alpha} + \frac{Q_\alpha}{H} Z,
\right)
\right].
\label{bardeenintegrate}
\end{equation}
By summing (\ref{bardeenintegrate}) with respect to all the fluid
components, we obtain      
\begin{eqnarray}
\left( h Z \right)_f &=&
\frac{H}{a^3} \left[
 \sum_{\alpha} C_\alpha + \int_{t_0} dt a^3 \left(
 - h_f {\cal A} 
 - \left( DP \right)_f 
 - S \cdot D \phi
\right)
\right]
\notag\\
&=&
\frac{H}{a^3} \left[ 
 \sum_{\alpha} C_\alpha +
\left(
 - \frac{2}{3} \frac{a^3}{H} \rho {\cal A}
 + a^3 \dot{\phi} \cdot D \phi
\right) -
\left(
 - \frac{2}{3} \frac{a^3}{H} \rho {\cal A}
 + a^3 \dot{\phi} \cdot D \phi
\right)_0
\right]
\end{eqnarray}
Therefore (\ref{constraint}) gives the constraint between $C_\alpha$,
$c$ defined by (\ref{conservquantity}), and $2 N_S + N_f$ solution constants
of the exactly homogeneous perturbation as
\begin{equation}
 \sum_{\alpha} C_\alpha +
\frac{2}{{\kappa}^2} c -
\left(
 - \frac{2}{3} \frac{a^3}{H} \rho {\cal A}
 + a^3 \dot{\phi} \cdot D \phi
\right)^{\sharp}_0 =0.
\label{constrainthomosol}
\end{equation}
Integrating (\ref{newtonpot}) gives
\begin{equation}
 \Phi = \frac{H}{a} \left( 
 C_t - \int_{t_0} dt a {\cal A}
\right),
\label{newtonintegrate}
\end{equation}
where the first term containing $C_t$ is well known universal adiabatic
decaying mode \cite{Kodama1998}
and by comparing with (\ref{conservquantity}) we obtain
\begin{equation}
 C_t = \frac{3}{k^2} c.
\end{equation}
If we assume $c=0$, since (\ref{constrainthomosol}) gives 
one constraint relation, we obtain $2 N_S + 2 N_f -1$ solutions
and the Newtonian potential is obtained by (\ref{newtonintegrate})
with $C_t =0$:
\begin{equation}
 \Phi = - \frac{H}{a} \int_{t_0} dt a {\cal A},
\end{equation}
If we assume that $c$ is nonvanishing, since $c=O(1)$, 
${\cal A} \to O(1)$, therefore
\begin{equation}
 \frac{{\cal A}}{C_t} \to O(k^2),
\end{equation}
we obtain the $k \to 0$ limit of the universal adiabatic
decaying mode \cite{Kodama1998}:
\begin{equation}
 \frac{1}{3} k^2 \Phi \to \frac{H}{a} c,
\end{equation}
which is consistent with (\ref{conservquantity}).
Then we have obtained the long wavelength limit
of all the solutions to the evolution equations of 
cosmological perturbations.

\section{Use of the scale factor as the evolution parameter}

As the gauge invariant variable representing the fluctuation of 
the scalar quantity $T$, we adopt $DT$ defined by (\ref{dtflatslice})
which represents the $T$ fluctuation in the flat slice, since
it is the easiest to see the correspondence with the exactly
homogeneous perturbation of $T$.
While until now we described the exactly homogeneous variables 
as functions of $t$, $C$ where $t$ is the cosmological time and
$C$'s are solution constants, we can describe the exactly homogeneous
variables as functions of $a$, $C$ where $a$ is the scale factor.
For an arbitrary scalar quantity 
such as $\rho$, $P$, $S$, $Q$, $\phi$,
from (\ref{dtsharp1}), (\ref{dtsharp2}), $(D T)^{\sharp}$ can be
written as the partial derivative of the corresponding exactly 
homogeneous scalar quantity $T$ with respect to solution constant $C$
with the scale factor $a$ fixed:
\begin{equation}
 \left( D T \right)^{\sharp} = 
 \left( \frac{\partial T}{\partial C} \right)_a .
\label{dtderivativeafixed}
\end{equation}
Since 
\begin{equation}
 \frac{1}{ \dot{a} }  
 \left( \frac{\partial}{\partial C} \dot{a} \right)_a
= \frac{1}{2 \rho}
 \left( \frac{\partial \rho}{\partial C} \right)_a
= \frac{1}{2 \rho}
 (D \rho)^{\sharp}
= - {\cal A}^{\sharp},
\end{equation}
this property of $D$ also holds as for the time derivative of 
the scalar quantity $\dot{T}$:
\begin{equation}
 \left( D \dot{T} \right)^{\sharp}
=  \left( \frac{\partial}{\partial C} \dot{T} \right)_a.
\end{equation}
Therefore the operator $D$ defined by (\ref{dtflatslice}) can be 
interpreted as a kind of derivative operator, that is, 
the derivative with respect to the solution constant $C$ with $a$ fixed.
Because of this derivative property of $D$, as for the scalar quantities
$T=(\rho, P, S, Q)$ which are functions of $\phi$, $\dot{\phi}$,
we can understand 
\begin{equation}
 D T = \frac{\partial T}{\partial \phi} \cdot D \phi +
 \frac{\partial T}{\partial \dot{\phi} } \cdot D \dot{\phi},
\end{equation}
easily.
The contribution to $D \dot{T}$ from the adiabatic decaying mode
is given by the same form as that of $D T$:
\begin{equation}
 \left( D \dot{T} \right)^{\flat} = \frac{\ddot{T}}{H} g,
\end{equation}
where $g$ is defined by (\ref{gfactor}).

As seen from the above duscussion, in order to derive 
the long wavelength limit of cosmological perturbations from the
corresponding exactly homogeneous system by using the LWL formula,
it is more appropriate to use as the evolution parameter 
the scale factor $a$ than the cosmological time $t$.
For example, as for the scalar-fluid composite system, the corresponding
exactly homogeneous expressions are obtained by solving the first order
differential equations 
setting $\phi_a$, $p_a := a^3 \dot{\phi}_a$, $\rho_{\alpha}$ as
independent variables and the scale factor $a$ as the evolution parameter:
\begin{eqnarray}
 a \frac{d}{d a} \phi_a &=& \frac{1}{H} \frac{p_a}{a^3},
\notag\\
 a \frac{d}{d a} p_a &=& 
 - \frac{a^3}{H} \frac{\partial U}{\partial \phi_a}
 - \frac{a^3}{H} S_a,
\notag\\
 a \frac{d}{d a} \rho_{\alpha} &=&
 - 3 h_{\alpha} + \frac{Q_{\alpha}}{H},
\end{eqnarray}
replacing $H$ with the right hand side of the Hubble law:
\begin{equation}
 H^2 = \frac{\kappa^2}{3} 
\left[
 \frac{1}{2 a^6} \sum_a p^2_a + U(\phi) + \sum_{\alpha} \rho_{\alpha}
\right].
\label{Hamiltonianconstraint}
\end{equation}
While under use of $t$ as the evolution parameter 
our system is the constrained system 
with the Hamiltonian constraint (\ref{Hamiltonianconstraint}), 
under use of $a$ as the evolution parameter our system becomes the 
unconstrained system with respect to independent variables
$\phi_a$, $p_a := a^3 \dot{\phi}_a$, $\rho_{\alpha}$.
The corresponding first order perturbation variables are
$D \phi_a$, $P_a := a^3 D \dot{\phi}_a$, $D \rho_{\alpha}$.

For some time, we consider the system consisting of multiple scalar 
fields $\phi_a$ only.
Since the evolutions of $\phi_a$, $p_a := a^3 \dot{\phi}_a$ can 
be described in terms of the Hamilton equations of motion,
the evolutions of the corresponding perturbation variables 
$Y_a =D \phi_a$ $P_a := a^3 D \dot{\phi}_a$ can also be written 
in terms of the Hamilton equations of motion:
\begin{equation}
 \frac{d Y_a}{d t} = \frac{\partial \bar{H}}{\partial P_a},
\quad \quad
 \frac{d P_a}{d t} = - \frac{\partial \bar{H}}{\partial Y_a},
\end{equation}
whose Hamiltonian is given by
\begin{eqnarray}
 \bar{H} &=& \frac{1}{2 a^3} P_a P_a 
 + \frac{a^3}{2} \bar{V}_{ab} Y_a Y_b
 + \frac{\kappa^2}{2 H} \dot{\phi}_a \dot{\phi}_b P_a Y_b,\\
\bar{V}_{ab} &=&
\frac{\partial^2 U}{\partial \phi_a \partial \phi_b}
+ \frac{3 \kappa^2}{2} \dot{\phi}_a \dot{\phi}_b
+ \frac{\kappa^2}{2 H} 
\left( 
   \frac{\partial U}{\partial \phi_a} \dot{\phi}_b
 + \dot{\phi}_a \frac{\partial U}{\partial \phi_b}
\right)
+ \frac{k^2}{a^2} \delta_{a b}.
\end{eqnarray}
When we discuss the quantization of the fluctuations,
the other set of the canonical perturbation variables 
$\tilde{Y}_a :=Y_a$, $\tilde{P}_a :=a^3 \dot{Y}_a$
has been used \cite{Mukhanov1992}.
But in the viewpoint of the LWL formalism, the set of the canonical 
variables $Y_a := D \phi_a$, $P_a := a^3 D \dot{\phi}_a$ is 
more natural than the set of the canonical variables
$\tilde{Y}_a :=Y_a$, $\tilde{P}_a :=a^3 \dot{Y}_a$, because
the long wavelength limit of the former set is generated from
the derivative of the homogeneous variables
$\phi_a$, $p_a := a^3 \dot{\phi}_a$ with respect to the solution
constants with the scale factor fixed.
The connection between the old canonical variables 
$\tilde{Y}_a :=Y_a$, $\tilde{P}_a :=a^3 \dot{Y}_a$ and 
the new canonical variables 
$Y_a := D \phi_a$, $P_a := a^3 D \dot{\phi}_a$ are 
given by the canonical transformation defined by the generating 
function:
\begin{equation}
 W = \tilde{Y}_a P_a + 
 \frac{3}{4} \frac{a^3 H}{\rho} 
 \dot{\phi}_a \dot{\phi}_b \tilde{Y}_a \tilde{Y}_b.
\end{equation}

When we treat the oscillatory scalar fields, the action angle variables
$I_a$, $\theta_a$ are useful \cite{Hamazaki2002}, \cite{Hamazaki2004}:
\begin{eqnarray}
 \phi_a &=& \frac{1}{a^{3/2}} \sqrt{\frac{2 I_a}{m_a}} \cos{\theta_a},
\notag\\ 
 p_a &=& - a^{3/2} \sqrt{2 m_a I_a} \sin{\theta_a},
\end{eqnarray}
where $m_a$ is the mass of the scalar field $\phi_a$.
The action angle variables obey the evolution equation as
\begin{eqnarray}
 a \frac{d}{da} I_a &=& 
 - \frac{a^3}{H} \frac{\partial U_{\rm int}}{\partial \theta_a} 
 + \frac{a^{3/2}}{H} \sqrt{\frac{2 I_a}{m_a}} \sin{\theta_a} \> S_a
 + 3 I_a \cos{2 \theta_a},
\label{actionevolution}\\
 a \frac{d}{da} \theta_a &=&
 \frac{m_a}{H} 
 + \frac{a^3}{H} \frac{\partial U_{\rm int}}{\partial I_a} 
 + \frac{a^{3/2}}{H} \frac{1}{\sqrt{2 m_a I_a}} \cos{\theta_a} \> S_a
 - \frac{3}{2} \sin{2 \theta_a}.
\label{angleevolution}
\end{eqnarray}
In order to investigate the cosmological perturbations in the universe
containing oscillatory scalar fields, by using $D$ defined in the above
we define the action angle perturbation variables $D I_a$, $D \theta_a$
starting from $Y_a :=D \phi_a$, $P_a :=a^3 D \dot{\phi}_a$.
In the LWL formalism,
the perturbation variables corresponding with the action angle variables
$I_a$, $\theta_a$ are $D I_a$, $D \theta_a$ defined by 
the following expressions:
\begin{eqnarray}
 Y_a &=& D\left[
\frac{1}{a^{3/2}} \sqrt{\frac{2 I_a}{m_a}} \cos{\theta_a}\right],
\notag\\ 
 P_a &=& D\left[- a^{3/2} \sqrt{2 m_a I_a} \sin{\theta_a}\right].
\label{defpertactionangle}
\end{eqnarray}
where $D$ in the right hand side is interpreted as
\begin{equation}
 D = \sum_a D I_a \frac{\partial}{\partial I_a} +
     \sum_a D \theta_a \frac{\partial}{\partial \theta_a}.
\end{equation}
The expressions obtained from variations of $I_a$, $\theta_a$ with $a$
fixed in the previous papers \cite{Hamazaki2002}, \cite{Hamazaki2004}
are the long wavelength limits of $D I_a$, $D \theta_a$ defined by
(\ref{defpertactionangle}).
In fact, the LWL formulae
\begin{eqnarray}
 D I_a &=&
 \left( \frac{\partial I_a}{\partial C} \right)_a +
\left( \frac{\dot{I}_a}{H}-3 I_a  \right) g,
\label{Dactionperturbformula}
\\
 D \theta_a &=&
 \left( \frac{\partial \theta_a}{\partial C} \right)_a +
 \frac{\dot{\theta}_a}{H} g,
\label{Dangleperturbformula}
\end{eqnarray}
where $g$ is defined in (\ref{gfactor}) hold.
While the third term in the right hand side of 
(\ref{Dactionperturbformula}) appears because of the scale factor $a$
dependence of the transformation law from $\phi_a$, $\dot{\phi}_a$ 
to $I_a$, $\theta_a$,
the $\sharp$ parts of (\ref{Dactionperturbformula}) and
(\ref{Dangleperturbformula}) reflect the fact that $D$ is the derivative
operator with respect to the solution constant with the scale factor $a$
fixed.

In order to solve the dynamics of the system containing the oscillatory
scalar fields, we are required to perform the averaging over the fast
changing angle variables $\theta_a$ 
\cite{Hamazaki2002}, \cite{Hamazaki2004}. 
If we use the cosmological time $t$ as the evolution parameter,
our system is a constrained system.
Therefore we must check that our averaging procedure is consistent with
the constraint and this process is rather cumbersome.
But if we use the scale factor $a$ as the evolution parameter, 
our system becomes unconstrained system, so the definition of the
averaging procedure becomes rather simple.

We can conclude that use of the scale factor $a$ as the evolution
parameter brings about the two merits.
One is that it becomes easier to see the correspondence between 
the exactly homogeneous solution and the long wavelength limit 
of the first order perturbation and that the LWL formulae become 
more simple.
The other is the more simple definition of the averaging process.

We consider the evolution of the Bardeen parameter $Z$.
Following the paper \cite{Wands2000}, we define $\zeta$ 
as the gauge invariant variable representing the curvature perturbation 
in the uniform density slice:
\begin{equation}
 \zeta := - \frac{H}{\dot{\rho}} D \rho 
 = {\cal R} - \frac{H}{\dot{\rho}} \delta \rho. 
\end{equation}
From (\ref{poissonperturb}) (\ref{constraint}) we can see that
the Baredeen parameter $Z$ and $\zeta$ are closely connected as
\begin{equation}
 Z = \zeta - \frac{2}{9} \frac{1}{1+w} \frac{k^2}{a^2 H^2} \Phi,
\end{equation}
whose $k \to 0$ limit is
\begin{equation}
  Z = \zeta - \frac{2}{3} \frac{1}{1+w} \frac{c}{a^3 H},
\end{equation}
where $c$ is constant related with the adiabatic decaying mode
defined by (\ref{conservquantity}).
While the Bardeen parameter $Z$ is expressed as the weighted sum of
$Z_S$, $Z_{\alpha}$ which do not related with the exactly
homogeneous quantity at all and whose evolution is written in the 
rather cumbersome integral form (\ref{bardeenintegrate}),
$\zeta$ is connected with the corresponding exactly homogeneous 
quantity and $\zeta^{\sharp}$ evolution can be written in terms of 
the derivative of the total energy density $\rho$ 
with respect to solution constant. 
Then we consider much easier $\zeta^{\sharp}$ evolution.

In the paper \cite{Lyth2005}, as the nonlinear generalization
of the Bardeen parameter $\zeta$, $\zeta (t, \bm{x})$ was introduced.
When $P = P(\rho)$, $\zeta (t, \bm{x})$ is reduced to
\begin{equation}
 \zeta (t, \bm{x}) = \ln{a(t, \bm{x})} + 
 \frac{1}{3} \int^{\rho (t, \bm{x})} \frac{d \rho}{\rho + P(\rho)}.
\end{equation}
In fact, the first order quantity of $\zeta (t, \bm{x})$ is given by
\begin{equation}
 \zeta_1 (t, \bm{x}) = \frac{\delta a(t, \bm{x})}{a}
 + \frac{1}{3} \frac{\delta \rho(t, \bm{x})}{\rho + P(\rho)},
\end{equation}
which agrees with the Bardeen parameter $\zeta$.
In the viewpoint of the LWL formalism, we adopt the zero curvature 
slice $\partial a(t, \bm{x}) / \partial \bm{x}^i = 0$. 
We assume the equation of state $P_r = \rho_r / 3$, since the final 
state of reheating and of the curvaton decay is radiation dominant.
In this case, the above $\zeta (t, \bm{x})$ is reduced to
\begin{equation}
 \zeta (a, \bm{x}) = \frac{1}{4} \ln{\rho_r (a, C(\bm{x}))}.
\label{bardeengenerator}
\end{equation}
$\rho_r (a, C(\bm{x}))$ is the expression of $\rho_r$ obtained 
by solving the locally homogeneous system (see the separate universe
approach \cite{Wands2000}) with use of $a$ as the evolution parameter.
The solution constants $C(\bm{x})$ have spatial dependence.
By considering $C(\bm{x})= C + \delta C(\bm{x})$ and expanding with 
respect to $\delta C(\bm{x})$, we can obtain the perturbation of an 
arbitrary order up to the decaying modes of perturbations. 
For example, the $n$-th order perturbation is given by
\begin{equation}
 \frac{1}{n!} \sum_{a_i} 
 \frac{\partial^n}
 {\partial C_{a_1} \partial C_{a_2} \cdot \cdot \cdot \partial C_{a_n}}
 \left[ \frac{1}{4} \ln{\rho_r (a, C)} \right]
 \delta C_{a_1}(\bm{x}) \delta C_{a_2}(\bm{x}) \cdot \cdot \cdot
 \delta C_{a_n}(\bm{x}).
\end{equation}
When we obtain the exactly homogeneous exression 
$\rho_r = \rho_r (a, C)$, we can know the long wavelength limits of 
perturbations of arbitrary orders.
Later we determine the exressions $\rho_r = \rho_r (a, C)$ in the 
reheating and in the curvaton decay.

In the reheating and in the curvaton decay, $\delta C(\bm{x})$'s are
given by the action angle variables in the initial time:
\begin{equation}
 \delta I_a (\bm{x}) :=  \delta I_a (a=a_0, \bm{x}), \quad \quad
 \delta \theta_a (\bm{x}) :=  \delta \theta_a (a=a_0, \bm{x}).
\end{equation}  
We discuss how to determine the statistical properties of 
$\delta I_a (\bm{x})$, $\delta \theta_a (\bm{x})$.
$\delta I_a (\bm{x})$, $\delta \theta_a (\bm{x})$ are given at the time
when the slow rolling phase ends and the coherent oscillation begins.
At this time, $\dot{\phi}_a = 0$, and as for the perturbations
\begin{equation}
 \delta \phi_a (\bm{x}) = \int d^3 \bm{k} e^{i \bm{k} \cdot \bm{x}} 
e_a (\bm{k}), \quad \quad  \delta \dot{\phi}_a (\bm{x}) = 0, 
\end{equation}
where $e_a (\bm{k})$ is the Gaussian random variable satisfying
\begin{equation}
 < e_a (\bm{k}) e_b (\bm{k}') > = P_a (k) \delta_{ab}  
 \delta^3 (\bm{k} + \bm{k}'), \quad \quad k := |\bm{k}|.
\end{equation}
By solving $I_a$ $\theta_a$ in terms of $\phi_a$ $\dot{\phi}_a$ and
by Taylor expanding, we obtain
\begin{eqnarray}
 \delta I_a (\bm{x}) &=& a^3_0 m_a \phi_a \delta \phi_a (\bm{x})
 + \frac{1}{2} a^3_0 m_a \left[ \delta \phi_a (\bm{x}) \right]^2,
\label{actiongaussian}\\
 \delta \theta_a (\bm{x}) &=& 0,
\end{eqnarray}
where we use $\dot{\phi}_a = 0$, $\delta \dot{\phi}_a (\bm{x})= 0$. 
The above fact that $\delta \theta_a (a=a_0, \bm{x}) =0$ does not imply
$\delta \theta_a (a, \bm{x})=0$, since $\theta_a (a)$ is a function 
depending on not only $\theta_a (a_0)$ but also $I_a (a_0)$.
Therefore it is often important to consider the role of the
perturbations of angle variables.

Determining the many point correlation function of fluctuations is
reduced to the evaluation of
\begin{equation}
  < e_{a_1} (\bm{k}_1) e_{a_2} (\bm{k}_2) \cdot \cdot \cdot 
    e_{a_n} (\bm{k}_n) >.
\end{equation}
This quantity can be determined by applying the differential 
operation defined by
\begin{equation}
 \exp \left[ \sum_a \int d^3 \bm{k} P_a (k) 
 \frac{\delta}{\delta e_a (\bm{k})}  
 \frac{\delta}{\delta e_a (- \bm{k})}
    \right]
 \cdot \cdot \cdot \Big|_{e=0}.
\end{equation}

\section{Application of the LWL formula to the non-interacting
multicomponent system}

Based on the results obtained in sections $2$, $3$, 
we consider the long wavelength limit of the evolutions of 
cosmological perturbations in the universe 
consisting of multiple cosmic components.
For simplicity, we consider the case where each component do not
interact with each other.

We consider the $w_{\alpha}$ fluid where $w_{\alpha}$ is constant
and which does not interact with other components $Q_{\alpha} = 0$.
$\rho_{\alpha}$ is solved as  
\begin{equation}
 \rho_{\alpha} = \frac{A_{\alpha}}{a^{3 (1+w_{\alpha})}},
\end{equation}
where $A_{\alpha}$ is a solution constant
and by differentiating with respect to $A_{\alpha}$ with the scale
factor $a$ fixed,
we obtain
\begin{equation}
 \left(D \rho_{\alpha} \right)^{\sharp}
 = \frac{\delta A_{\alpha}}{a^{3 (1+w_{\alpha})}},
\end{equation}
where $\delta A_{\alpha}$ is a perturbation solution constant 
corresponding with $A_{\alpha}$.
Therefore we obtain
\begin{equation}
 \Delta_{g \alpha}^{\sharp} = \frac{\delta A_{\alpha}}{A_{\alpha}}
 = {\rm const}.
\end{equation}
Next we consider the case that the oscillatory scalar field $\phi_a$
does not interact with other cosmic components, 
that is $U_{\rm int} =0$, $S_a = 0$.
Since in such case the right hand side of (\ref{actionevolution}) 
is oscillatory function depending on the angle variable $\theta_a$
with vanishing mean value,
by taking the averaging over $\theta_a$, we obtain
\begin{equation}
 I_a \cong A_a,
\end{equation}
where $A_a$ is constant.
The estimate about the effects of the oscillations due to the fast 
changing angle variables $\theta_a$ was discussed in the previous 
papers \cite{Hamazaki2002} \cite{Hamazaki2004}.
Therefore by taking the derivative with respect to solution constant 
$A_a$ with the scale factor $a$ fixed, we obtain 
\begin{equation}
 D I_a^{\sharp} \cong \delta A_a,
\end{equation}
where $\delta A_a$ is a perturbation constant corresponding with $A_a$.
Since 
\begin{equation}
 \rho_a = \frac{m_a I_a}{a^3}, \quad \quad
 D \rho_a = \frac{m_a D I_a}{a^3},
\end{equation}
where the above second expression is given by the 
$D$ operation to the above first expression,
we obtain
\begin{equation}
 \Delta_{g a}^{\sharp} = \frac{D I_a^{\sharp}}{I_a}
 \cong \frac{\delta A_a}{A_a}
 = {\rm const}.
\end{equation}
As seen from the above expressions, 
as for the oscillatory scalar field $\phi_a$
the energy density $\rho_a$ and
the energy density perturbation $\rho_a \Delta_{g a}^{\sharp}$
behave like those of the dust fluid.   
We can summarize that for non-interacting cosmic components,
$\Delta_{g \alpha}^{\sharp}$, $\Delta_{g a}^{\sharp}$ are conserved.

As for the multicomponent non-interacting fluids system, 
$\zeta^{\sharp}$ evolution is given by 
\begin{equation}
 \zeta^{\sharp} = 
\left(
 \sum_{\alpha} \frac{\delta A_{\alpha}}{a^{3 (1+w_{\alpha})}} \right) \big/
\left(
 3 \sum_{\alpha} (1+w_{\alpha}) \frac{A_{\alpha}}{a^{3 (1+w_{\alpha})}}
\right).
\label{noninteractbardeen}
\end{equation}
We can see that $\zeta^{\sharp}$ is exactly conserved for the adiabatic
growing mode defined by
\begin{equation}
 \frac{\delta A_{\alpha}}{A_{\alpha}} \frac{1}{1+w_{\alpha}} 
 = \alpha \> {\rm independent}.
\end{equation}

For some time, we consider the two components system consisting
of dust and radiation.
In this case, (\ref{noninteractbardeen}) is reduced to
\begin{equation}
 \zeta^{\sharp} = 
\frac{a \delta A_d + \delta A_r}{3 a A_d + 4 A_r},
\label{radiationdustbardeen}
\end{equation}
where the suffix $d$, $r$ imply dust, radiation, respectively.
We obtain
\begin{equation}
 \zeta^{\sharp}_{\rm init} = \frac{1}{4} \frac{\delta A_r}{A_r},
\quad \quad
 \zeta^{\sharp}_{\rm fin} = \frac{1}{3} \frac{\delta A_d}{A_d},
\end{equation}
in the limit $a \to 0$, $a \to \infty$, respectively.
We adopt different, more physical paramerization:
\begin{equation}
 \delta A_d = 3 \xi A_d - \eta A_d, \quad \quad
 \delta A_r = 4 \xi A_r.
\end{equation}
$\xi$ represents the adiabatic growing mode and $\eta$ represents
the isocurvature mode defined by 
\begin{equation}
 \zeta^{\sharp}_{\rm init} = 0, \quad \quad
 S^{\sharp}_{rd} = \frac{3}{4} \Delta^{\sharp}_{g r} 
 - \Delta^{\sharp}_{g d} = {\rm const} =: \eta. 
\end{equation}
Then we obtain
\begin{equation}
 \zeta^{\sharp}_{\rm fin} = \zeta^{\sharp}_{\rm init} 
 - \frac{1}{3} \eta,
\end{equation}
which is the famous formula.\cite{Kodama1987}

Although in the paper \cite{Wands2000},
(\ref{radiationdustbardeen}) has already been derived
essentially without using the LWL formula, 
our result is more rigorous in the
point that we treat the contribution from the adiabatic decaying mode
characterized by $c$ defined by (\ref{conservquantity}) more
appropriately, while the paper \cite{Wands2000} simply assumes that
$k^2 \Phi$ is vanishing.

\section{Application of the averaging method to the decaying scalar
fields}

We derive the evolution equations of the multiple scalar fields
decaying into the multiple radiation fluids.
By solving these evolution equations and taking the exactly 
homogeneous perturbations, we can obtain the information of the 
evolutionary behaviors of cosmological perturbations during 
reheating and in the curvaton model.
We assume that the source $S_a$ is given by
\begin{equation}
 S_a = \Gamma_a \dot{\phi}_a.
\end{equation}
We nondimensionalize the dynamical quantities as
\begin{equation}
 \frac{a}{a_0} \to a, \quad
 \frac{I_a}{I_0} \to I_a, \quad
 \frac{\rho_{\alpha}}{\rho_0} \to \rho_{\alpha}, \quad  
 \frac{U_{\rm int}}{\rho_0} \to U_{\rm int} = O(\nu),
\end{equation}
and the parameters as
\begin{equation}
 \frac{m_a}{m_0} \to m_a, \quad
 \frac{\Gamma_a}{\Gamma_0} \to \Gamma_a, \quad 
\end{equation}
where $\nu$ is the small parameter implying the ratio of the 
interaction energy to the free part energy defined by
\begin{equation}
 \rho_0 := \frac{m_0 I_0}{a^3_0}.
\end{equation}
Then we obtain the dimensionless parameters 
\begin{equation}
 \epsilon := \frac{\kappa}{\sqrt{3}} \frac{\rho^{1/2}_0}{m_0}, \quad
 \gamma := \frac{\sqrt{3} \Gamma_0}{\kappa \rho^{1/2}_0},
\end{equation}
which imply the ratio of the Hubble parameter to the mass of the scalar
fields $H/m$, the ratio of the decay rate to the Hubble parameter 
$\Gamma/H$ at the initial time $a= a_0$, respectively.
By using the above dynamical variables and parameters, the evolution
equations can be expressed as 
\begin{eqnarray}
 a \frac{d}{da} I_a &=& 
 - \frac{1}{\epsilon} \frac{a^3}{\rho^{1/2}}
 \frac{\partial U_{\rm int}}{\partial \theta_a} 
 - \gamma \Gamma_a \frac{I_a}{\rho^{1/2}} 
 \left( 1 - \cos{2 \theta_a} \right)
 + 3 I_a \cos{2 \theta_a},\\
 a \frac{d}{da} \theta_a &=&
 \frac{1}{\epsilon} \frac{m_a}{\rho^{1/2}} 
 + \frac{1}{\epsilon} \frac{a^3}{\rho^{1/2}}
  \frac{\partial U_{\rm int}}{\partial I_a} 
 - \frac{1}{2} \gamma \frac{\Gamma_a}{\rho^{1/2}} 
   \sin{2 \theta_a}
 - \frac{3}{2} \sin{2 \theta_a},\\
 a \frac{d}{da} \sigma_{\alpha} &=& 
  \gamma \frac{a}{\rho^{1/2}} \sum_a \Gamma_{\alpha a}
  m_a I_a 
 \left( 1 - \cos{2 \theta_a} \right),
\end{eqnarray}
where $\sigma_{\alpha}$ is defined by
\begin{equation}
 \rho_{\alpha} = \frac{\sigma_{\alpha}}{a^4},
\end{equation}
and $\Gamma_{\alpha a}$ is the decay rate from the scalar field $I_a$
to the radiation component $\sigma_{\alpha}$ and therefore $\Gamma_a$
is given by
\begin{equation}
 \Gamma_a = \sum_{\alpha} \Gamma_{\alpha a}.
\end{equation}
In this paper, we investigate the evolutionary behavior of cosmological
perturbations during the period when the decay rates from the scalar 
fields to the radiation fluids are large compared to the interaction
between the scalar fields, that is $\gamma \gg \nu / \epsilon$,
while in the paper \cite{Hamazaki2004} the evolutions of cosmological
perturbations during the period when the interaction between the 
scalar fields is dominant $\nu / \epsilon \gg \gamma$ were discussed,
which is thought to give the initial conditions for the present 
studies.

Next we show that there exists a transformation such that in a 
system obtained by that transformation, the dynamics of 
the action variables $I_a$ and the radiation energy densities 
$\sigma_{\alpha}$ can be determined independently of the 
angle variables $\theta_a$.
In order to show this statement, we put several assumptions.
\begin{itemize}
 \item[(i)] The interaction energy of the scalar fields $U_{\rm int}$
  is analytic with respect to the dynamical variables $I_a$, $\theta_a$
  and is $2 \pi$ periodic with respect to $\theta_a$.
  $U_{\rm int}$ is bounded as
\begin{equation}
 U_{\rm int} \sim \frac{\nu}{a^{9/2}} |I|,
\end{equation} 
 which implies that there exists a positive constant $M$
 such that 
\begin{equation}
 |a^{9/2} U_{\rm int}| \le \nu M |I|,
\end{equation}
 where 
\begin{equation}
 |I| := \sum_a |I_a|.
\end{equation}
 \item[(ii)] As for the total energy density
\begin{equation}
 \rho = \sum_a \frac{m_a I_a}{a^3} 
      + \sum_{\alpha} \frac{\sigma_{\alpha}}{a^4}
      + U_{\rm int},
\end{equation}
 the action variables $I_a$ and the radiation energy densities 
 $\sigma_{\alpha}$ satisfy
\begin{equation}
 c_1 \le a \sum_a m_a I_a
         + \sum_{\alpha} \sigma_{\alpha} \le c_2,
\label{assumptiontwo}
\end{equation}
 for some positive constants $c_1$, $c_2$.
\end{itemize}
Note that the evolution equations of the system can be written as  
\begin{eqnarray}
 a \frac{d}{da} I_a &=& F_a (I, \sigma, \theta, a),\\
 a \frac{d}{da} \sigma_{\alpha} &=& F_{\alpha} (I, \sigma, \theta, a),\\
 a \frac{d}{da} \theta_a &=& \frac{1}{\epsilon} \omega_a (I, \sigma, a)
 + G_a (I, \sigma, \theta, a),
\end{eqnarray}
where $F_a$ $F_{\alpha}$ $G_a$ are analytic with respect to the 
dynamical variables $I_a$ $\sigma_{\alpha}$ $\theta_a$ and
$2 \pi$ periodic with respect to the angle variables $\theta_{\alpha}$.
We say that the evolution equations is of the type $C_k$:
\begin{itemize}
 \item[(i)]The averaged parts of $F_a$ $F_{\alpha}$ $G_a$ are bounded as
\begin{eqnarray}
 <F_a> &\sim& a^2 |I|,\\
 <F_{\alpha}> &\sim& a^3 |I|,\\
 <G_a> &\sim& a^2,
\end{eqnarray}
 where $<A>$ implies the averaging over the angle variables $\theta_a$:
\begin{equation}
 <A> := \frac{1}{(2 \pi)^{N_S}} \int^{2 \pi}_0 d^{N_S} \theta \> A.
\end{equation}
 In case of the resonant case, it is prescribed that the averaging
 is performed with respect to the fast angle variables only
 \cite{Hamazaki2004}.
 \item[(ii)]The oscillatory parts of $F_a$ $F_{\alpha}$ $G_a$
 are bounded as
\begin{eqnarray}
 \tilde{F}_a &\sim& \epsilon^k a^2 |I|,\\
 \tilde{F}_{\alpha} &\sim& \epsilon^k a^3 |I|,\\
 \tilde{G}_a &\sim& \epsilon^k a^2,
\end{eqnarray}
 where $\tilde{A}$ implies the residual part after the averaging 
 over the angle variables $\theta_a$:
\begin{equation}
 \tilde{A} := A - <A>.
\end{equation}
\end{itemize}
Under this notation, the following proposition holds.
 
\paragraph{Proposition $1$}

\textit{Let $k$ be some non-negative integer and consider the 
evolution equations as}
\begin{eqnarray}
 a \frac{d}{da} I^{(k)}_a &=& 
 F^{(k)}_a (I^{(k)}, \sigma^{(k)}, \theta^{(k)}, a),\\
 a \frac{d}{da} \sigma^{(k)}_{\alpha} &=& 
 F^{(k)}_{\alpha} (I^{(k)}, \sigma^{(k)}, \theta^{(k)}, a),\\
 a \frac{d}{da} \theta^{(k)}_a &=& 
 \frac{1}{\epsilon} \omega^{(k)}_a (I^{(k)}, \sigma^{(k)}, a)
 + G^{(k)}_a (I^{(k)}, \sigma^{(k)}, \theta^{(k)}, a).
\end{eqnarray}
\textit{Suppose that this set of evolution equations is of the type
$C_k$, there exists a transformation}
\begin{eqnarray}
 I^{(k)}_a &=&  I^{(k+1)}_a +
 u^{(k)}_a (I^{(k+1)}, \sigma^{(k+1)}, \theta^{(k+1)}, a),\\
 \sigma^{(k)}_{\alpha} &=& \sigma^{(k+1)}_{\alpha} +
 u^{(k)}_{\alpha} (I^{(k+1)}, \sigma^{(k+1)}, \theta^{(k+1)}, a),\\
 \theta^{(k)}_a &=&  \theta^{(k+1)}_a +
 v^{(k)}_a (I^{(k+1)}, \sigma^{(k+1)}, \theta^{(k+1)}, a),
\end{eqnarray}
\textit{satisfying the following conditions:}
\begin{itemize}
 \item[(i)] \textit{
$u^{(k)}_a$ $u^{(k)}_{\alpha}$ $v^{(k)}_a$ are analytic with respect to the 
dynamical variables $I^{(k+1)}_a$ $\sigma^{(k+1)}_{\alpha}$
$\theta^{(k+1)}_a$, are $2 \pi$ periodic with respect to the angle
variables $\theta^{(k+1)}_a$, and are bounded as} 
\begin{eqnarray}
 u^{(k)}_a &\sim& \epsilon^{k+1} |I^{(k+1)}|,\\
 u^{(k)}_{\alpha} &\sim& \epsilon^{k+1} a |I^{(k+1)}|,\\
 v^{(k)}_a &\sim& \epsilon^{k+1}.
\end{eqnarray}
 \item[(ii)]\textit{The evolution equations of the transformed variables
$I^{(k+1)}_a$ $\sigma^{(k+1)}_{\alpha}$ $\theta^{(k+1)}_a$ are of the 
type $C_{k+1}$ and the changes of $<F_a>$ $<F_{\alpha}>$ $<G_a>$ are 
bounded as}
\begin{eqnarray}
 \Delta <F_a> &\sim& \epsilon^{k+1} a^2 |I^{(k+1)}|,\\
 \Delta <F_{\alpha}> &\sim& \epsilon^{k+1} a^3 |I^{(k+1)}|,\\
 \Delta <G_a> &\sim& \epsilon^{k+1} a^2,
\end{eqnarray}
\textit{where $\Delta <A>$ is defined by}
\begin{equation}
 \Delta <A> := 
<A^{(k+1)}(I^{(k+1)}, \sigma^{(k+1)}, \theta^{(k+1)}, a)> -
<A^{(k)}(I^{(k+1)}, \sigma^{(k+1)}, \theta^{(k+1)}, a)>.
\end{equation}
\end{itemize}
(For the proof, see Appendix $A$.)

This proposition implies that we can make the part depending on the 
angle variables arbitrarily small by taking the original set of the 
evolution equations as the starting point $k = 0$ and applying 
transformations given in the proposition iteratively.
Therefore it can be expected that the evolution of the dynamical
variables can be described by the truncated system obtained
by discarding the angle variable dependent part with sufficiently
good accuracy, if we take sufficiently large $k$.
By estimating the errors produced by the truncation, we show that
the above expectation is correct.
We use the symbol $\Delta$ to represent the difference of a quantity
for the exact system and a corresponding quantity for the truncated
system.
For $A=(I_a, \sigma_{\alpha}, \theta_a)$, the errors of the background
variables are written as
\begin{equation}
 \Delta A := A - A_{\rm tr}, 
\end{equation}
and the errors of the perturbation variables are written as 
\begin{equation}
 \Delta \delta A := \delta A - \delta A_{\rm tr}, 
\end{equation}
where $A$, $\delta A$ represent quantities of the exact system and
$A_{\rm tr}$, $\delta A_{\rm tr}$ represent quantities of the 
corresponding truncated system.
For a function $f(a)$, let us write
\begin{equation}
 f(a) = E(- a^2),
\end{equation}
when $f(a)$ is bounded as
\begin{equation}
 |f(a)| \le p(a) \exp{(- \lambda a^2)}
\end{equation}
for a polynomial of $a$: $p(a)$, 
and for a positive number $\lambda$.
For a function $f(a)$, let us define $\| f(a) \|$ by
\begin{equation}
 \| f(a) \| := \sup_{1 \le a' \le a} |f(a')|.
\end{equation}
When we write all the inequalities, it is prescribed that
all the coefficients of order unity are omitted.
The truncation error for the $m$-th order system can be estimated 
as follows.

\paragraph{Proposition $2A$}

\textit{Let $m$ be an integer larger than or equal to $2$.
For the $m$-th order system, the truncation errors of the background
variables are given by}
\begin{eqnarray}
 |\Delta I| &\le& E(- a^2) \epsilon^m,\\
 \| \Delta \sigma \| (a) &\le& \epsilon^m,\\
 |\Delta \theta| &\le& a^2 \epsilon^{m-1},
\end{eqnarray}
\textit{and the truncation errors of the perturbation variables
are given by}
\begin{eqnarray}
 |\Delta \delta I| &\le& E(- a^2) \epsilon^{m-1} \delta A_1 (1),\\
 |\Delta \delta \sigma | &\le& \epsilon^{m-1} \delta A_1 (1),\\
 |\Delta \delta \theta| &\le& a^2 \epsilon^{m-2} \delta A_1 (1)
 + \epsilon^{m-1} \exp{(a^2 \epsilon^m)} 
 \left[
a^2 \epsilon |\delta \theta(1)| + a^4 \delta A_m (1)
 \right],
\end{eqnarray}
\textit{where}
\begin{equation}
 \delta A_m (1) := |\delta I(1)|+|\delta \sigma (1)|+
 \epsilon^m |\delta \theta(1)|,
\label{deltaAm}
\end{equation}
\textit{under the initial conditions}
\begin{eqnarray}
 && \Delta I(1) = \Delta \sigma (1) = \Delta \theta(1)=0,\\
 && \Delta \delta I(1) = \Delta \delta \sigma (1) = 
 \Delta \delta \theta(1)=0.
\end{eqnarray}

(For the proof, see Appendix $A$.)

By the transformation laws, the errors of the $m$-th order variables
affect the original variables as shown by the next proposition.

\paragraph{Proposition $2B$}

\textit{The difference between $A^{(0)}$ obtained from $A^{(m)}$
by the transformation laws and $A^{(0)}_{\rm tr}$ obtained 
from $A^{(m)}_{\rm tr}$ by the same transformation laws has 
upper bound}
\begin{eqnarray}
 |\Delta I^{(0)}| &\le& E(- a^2) \epsilon^m,\\
 |\Delta \sigma^{(0)} | &\le& \epsilon^m,\\
 |\Delta \theta^{(0)}| &\le& a^2 \epsilon^{m-1},
\end{eqnarray}
\textit{and the corresponding difference as for the perturbation 
variables has upper bound}
\begin{eqnarray}
 |\Delta \delta I^{(0)}| &\le& 
 E(- a^2) \epsilon^{m-1} \delta A^{(m)}_1 (1),\\
 |\Delta \delta \sigma^{(0)} | &\le& 
 \epsilon^{m-1} \delta A^{(m)}_1 (1),\\
 |\Delta \delta \theta^{(0)}| &\le& 
 a^2 \epsilon^{m-2} \delta A^{(m)}_1 (1)
 + \epsilon^{m-1} \exp{(a^2 \epsilon^m)} 
 \left[
a^2 \epsilon |\delta \theta^{(m)} (1)| + a^4 \delta A^{(m)}_m (1)
 \right],
\end{eqnarray}
\textit{under the same initial condition as in the previous proposition.}

(For the proof, see Appendix $A$.)

According to the above proposition, we can conclude that we can 
make the truncation errors as for the original variables arbitrarily
small if we truncate the system at the arbitarily large $m$-th order
system.

From Proposition $1$, we can see that the part independent of 
the angle variables of the evolution equations are shifted after 
the transformations reducing the part dependent on the angle variables.
By the truncation, our system become much simpler than the original
system.
But it is still difficult to solve the truncated evolution equations
with such correction terms because the evolution equations are
complicatedly entangled with each other.
In order to solve the evolution equations analytically,
we want to discard such correction terms.
The errors produced by discarding such corrections are evaluated 
in the following proposition. 
   
\paragraph{Proposition $3$}

\textit{The difference between the system with the correction
terms produced by the transformation and the system obtained 
by discarding such correction terms is evaluated in the following 
way.
As for the background variables, the discard errors are 
evaluated as}
\begin{eqnarray}
 |\Delta I| &\le& E(- a^2) \epsilon,\\
 |\Delta \sigma| &\le& \epsilon,\\
 |\Delta \theta| &\le& a^2,
\end{eqnarray}
\textit{and as for the perturbation variables, 
the discard errors are evaluated by}
\begin{eqnarray}
 |\Delta \delta I| &\le& E(- a^2) \epsilon \delta B (1),\\
 |\Delta \delta \sigma | &\le& \epsilon \delta B (1),\\
 |\Delta \delta \theta| &\le& a^2 \delta B (1), 
\end{eqnarray}
\textit{where}
\begin{equation}
 \delta B (1) := |\delta I(1)|+|\delta \sigma (1)|,
\label{deltaB}
\end{equation}
\textit{under the initial conditions}
\begin{eqnarray}
 && \Delta I(1) = \Delta \sigma (1) = \Delta \theta(1)=0,\\
 && \Delta \delta I(1) = \Delta \delta \sigma (1) = 
 \Delta \delta \theta(1)=0.
\end{eqnarray}

(For the proof, see Appendix $A$.)

By the above propositions, it can be understood that we can 
obtain the information of the original system with sufficiently
good accuracy by investigating the evolution equations 
by simply dropping the part dependent on the angle variables,
because the errors produced by dropping are sufficiently 
small and in particular $\Delta \sigma$, $\Delta \delta \sigma$
are bounded.
The reason why these errors are mild is that the final state 
in which all the energy of the scalar fields is completely
transferred into that of radiation fluids is the attracting 
equilibrium around which the perturbations do not grow.

\section{Application of the LWL formula to the multicomponent reheating
model}

In this section, we apply the LWL formula to the reheating where 
the energy of the multiple scalar fields is transferred into that
of the multiple radiation fluids.
The decay rate from the scalar field $\phi_a$ to the radiation fluid
$\rho_{\alpha}$ is given by $\Gamma_{\alpha a}$.
When the interactions between the scalar fields $\phi_a$: $U_{\rm int}$
are neglected, the background quantities are solved as 
\begin{eqnarray}
 m_a I_a &=& A_a
 \exp \left\{ - \gamma \Gamma_a 
 \int_1 d a \frac{1}{\rho^{1/2} a} \right\},
\label{quasisolution1}\\
 \sigma_{\alpha} &=& 
 B_{\alpha} + \int_1 d a \frac{\gamma}{\rho^{1/2}}
 \sum_a \Gamma_{\alpha a} m_a I_a,
\label{quasisolution2}
\end{eqnarray}
where $A_a$, $B_{\alpha}$ are integration constants.
As long as we do not give the expression of the total energy density
$\rho$ in the integrals, the above solutions do not give any physical
information of reheating.
But it is difficult to solve the evolution equations in the form where
the exact expression of $\rho$ is explicitly described, because
the evolution equations of $I_a$, $\sigma_{\alpha}$ are complicated 
and highly nonlinear.
Then we expect that the contribution to the integrations owes mainly
to the period when the energy of the scalar fields is dominant, that is
$\rho$ can be approximated as 
\begin{equation}
 \rho = \frac{A}{a^3} + \frac{B}{a^4},
\label{earlyenergydensity}
\end{equation}
where
\begin{equation}
 A:=\sum_a A_a, \quad B:=\sum_{\alpha} B_{\alpha}.
\end{equation}
We assume that the initial radiation energy density $B$ is 
negligibly small $B \ll A$.
We assume that all $\Gamma_a := \sum_{\alpha} \Gamma_{\alpha a}$
($a=1,2, \cdot \cdot \cdot ,N_S $) are of the same order of
magnitude.
By substituting the above $\rho$ expression to the solutions
(\ref{quasisolution1})(\ref{quasisolution2}), by expanding 
the solutions with respect to $B$ around $B=0$,
we obtain
\begin{eqnarray}
 m_a I_a &=& A_a 
 \exp \left\{ - \frac{2}{3} \frac{\gamma \Gamma_a}{A^{1/2}}   
 a^{3/2}
      \right\}
 + \gamma \Gamma_a
 \frac{A_a B}{A^{3/2}} a^{1/2}
 \exp \left\{ - \frac{2}{3} \frac{\gamma \Gamma_a}{A^{1/2}}
 a^{3/2}
      \right\}+O(B^2),
\notag\\
a^4 \rho_r &=& a^4 \sum_{\alpha} \rho_{\alpha}
\notag\\
&=& 
 B \left\{ 1+ \frac{3}{2} G(2) - \frac{1}{2} G(1) \right\}
 + \left( \frac{3}{2} \right)^{2/3} G \left( 5/3 \right)
 \frac{A^{1/3}}{\gamma^{2/3}}
 \sum_a \frac{A_a}{\Gamma^{2/3}_a} + O(B^2),
\label{backgroundreheating}
\end{eqnarray}
where 
\begin{equation}
 \Gamma_a := \sum_{\alpha} \Gamma_{\alpha a},
\end{equation}
and $G(t)$ is the Gamma function defined by 
\begin{equation}
 G(t) := \int^{\infty}_{0} d x x^{t-1} e^{- x},
\end{equation}
which is convergent for $t > 0$.
By taking the exactly homogeneous perturbation of
(\ref{backgroundreheating}) defined by
\begin{equation}
 D := \left( \delta A \cdot \frac{\partial}{\partial A} +
             \delta B \cdot \frac{\partial}{\partial B} 
      \right)_{a \; B=0},
\end{equation}
we can obtain the Bardeen parameter in the final state $a \to \infty$: 
\begin{eqnarray}
 \zeta^{\sharp}_{\rm fin} &\cong& \frac{1}{4}
 \frac{D \rho^{\sharp}_r}{\rho_r}   
\notag\\
&=&
 \frac{1}{4} \left( \frac{2}{3} \right)^{2/3}
 \frac{\gamma^{2/3}}{A^{1/3}}
 \delta B \left\{ 1+ \frac{3}{2} G(2) - \frac{1}{2} G(1) \right\}
 \Big/ \sum_a \frac{A_a}{\Gamma^{2/3}_a} G \left( 5/3 \right) 
\notag\\ 
 && + \frac{1}{4}
\left\{ \frac{1}{3}
 \frac{\delta A}{A} \sum_a \frac{A_a}{\Gamma^{2/3}_a}
 + \sum_a \frac{\delta A_a}{\Gamma^{2/3}_a}
\right\}
 \Big/ \sum_a \frac{A_a}{\Gamma^{2/3}_a}.
\end{eqnarray}
From now on, we name the expressions representing the final amplitude
of the Bardeen parameter $\zeta^{\sharp}_{\rm fin}$ in terms of the 
initial perturbation amplitudes such as $\delta A_a$ 
$\delta B_{\alpha}$, S formula after S matrix in the quantum mechanics.
As for the initial energy density perturbations of the scalar fields
$\delta A_a$, by adopting more physical parametrization introduced by
\begin{equation}
 \frac{1}{3} \frac{\delta A}{A} =: \xi, \quad
 S_{ab} = \frac{\delta A_a}{A_a} - \frac{\delta A_b}{A_b}
 =: \eta_{ab},
\end{equation}
$\delta A_a$ can be written as
\begin{equation}
 \delta A_a = 3 A_a \xi + \sum_{b} \frac{A_a A_b}{A}
 \eta_{ab}.
\end{equation}
$\xi$ represents the adiabatic growing mode and $\eta_{ab}$ represent
the isocurvature modes.
Since $\eta_{ab}$ satisfy
\begin{equation}
 \eta_{ab} = - \eta_{ba}, \quad
 \eta_{ab} + \eta_{bc} = \eta_{ac},
\end{equation}
the independent quantities are given by $\xi$ $\eta_{12}$ $\eta_{23}$
$\cdot \cdot \cdot$ $\eta_{N_S-1 \> N_S}$.
By using this parametrization, $\delta A_a$ dependent part of  
$\zeta^{\sharp}_{\rm fin}$ is written as
\begin{equation}
 \zeta^{\sharp}_{\rm fin} \supset \xi + \frac{1}{8}
 \sum_{ab} \left( 
 \frac{1}{\Gamma^{2/3}_a} - \frac{1}{\Gamma^{2/3}_b} 
           \right)
 \frac{A_a A_b}{A} \eta_{ab}
 \Big/ \sum_a \frac{A_a}{\Gamma^{2/3}_a},
\label{Bardeenadiiso}
\end{equation}
where $A \supset B$ implies that $B$ is contained by $A$, 
that is $A = B + \cdot \cdot \cdot$. 
From this S formula, we can conclude that for the adiabatic growing
mode $\xi$, the Bardeen parameter is conserved, and that
the initial entropy perturbations survive in the case that
the decay rates are dependent on the scalar field $\phi_a$
from which the radiation energy comes, that is 
$\Gamma_a \neq \Gamma_b$ ($a \neq b$).
In the case where multiple scalar fields exist, there is no 
reason why the perturbation has only the adiabatic component, 
and it is natural to think that in the perturbation 
the adiabatic components and the entropic components coexist.
In such mixed cases, so called conservation of the Bardeen parameter
does not hold and the above S formula gives useful tool for 
calculating the final Bardeen parameter.

We can consider the case where the energy transfer rates 
$\Gamma_{\alpha a}$ fluctuates \cite{Sasaki1991}, 
which is called as the modulated reheating scenario 
\cite{Dvali2004}.
For simplicity, we consider the one scalar field case.
By taking the derivative of (\ref{backgroundreheating}) 
with respect to $\Gamma$ with $B$ vanishing, we obtain
\begin{equation}
 \zeta^{\sharp}_{\rm fin} \supset
 - \frac{1}{6} \frac{\delta \Gamma}{\Gamma},
\label{modulatereheating}
\end{equation}
which is well known formula derived in the paper \cite{Dvali2004},
and where the coefficient is successfully determined in this paper.

We consider the influence of the resonant interaction between 
scalar fields on the final amplitude of the Bardeen parameter.
We take the interaction $U_{\rm int}$ into account by iteration. 
As the first order correction from the interaction term as
\begin{equation}
 -\frac{1}{\epsilon} \frac{a^3}{\rho^{1/2}}
 \frac{\partial U_{\rm int}}{\partial \theta_a} 
 \subset a \frac{d}{da} I_a,
\end{equation}
we obtain
\begin{equation}
 \sigma_{\alpha} \supset 
 \int_1 d a \frac{\gamma}{\rho^{1/2}}
 \sum_a \Gamma_{\alpha a} 
 \Delta \left( m_a I_a \right),
\end{equation}
where
\begin{equation}
\Delta \left( m_a I_a \right) :=
  \exp \left\{ - \gamma \Gamma_a
 \int_1 d a \frac{1}{\rho^{1/2} a} \right\}
 \int_1 d a \frac{m_a}{a}
  \exp \left\{ \gamma \Gamma_a
 \int_1 d a \frac{1}{\rho^{1/2} a} \right\}
 \left( -\frac{1}{\epsilon} \frac{a^3}{\rho^{1/2}}
 \right)
 \frac{\partial U_{\rm int}}{\partial \theta_a}.
\end{equation}
Since as the zeroth order approximation $I_a$ obeys
(\ref{quasisolution1}), we can write
\begin{equation}
 \frac{\partial U_{\rm int}}{\partial \theta_a} =
 \frac{\partial U_{\rm int}}{\partial \theta_a} \Big|_{a=1}
 \frac{1}{a^{3 n / 2}}
 \exp \left\{ - \gamma \Gamma
 \int_1 d a \frac{1}{\rho^{1/2} a} \right\},
\end{equation}
where we assumed that $U_{\rm int}$ contains an n-th order interaction
term and that $\Gamma$ is the appropriate sum of $\Gamma_a / 2$.
By substituting the above expression and by expanding the correction
term with respect to $B$ around $B=0$, we obtain 
\begin{eqnarray}
 \sigma_a &\supset& 
 \sum_a \Gamma_{\alpha a} m_a 
 \frac{\partial U_{\rm int}}{\partial \theta_a} \Big|_{a=1}
 \left( - \frac{1}{\epsilon} \right)
 \left( \frac{2}{3} \frac{\gamma}{A^{1/2}} \right)^{n-8/3}
 \frac{1}{\gamma}
\notag\\
 &\times& \left[ G(5/3, -n+3, \Gamma_a, \Gamma) +
 \left( \frac{\gamma}{A^{1/2}} \right)^{2/3}
 \frac{B}{A}
\left\{
 \left( \frac{1}{2} \right)^{1/3}
 \left( \Gamma - \Gamma_a \right)
 G(5/3, -n+10/3, \Gamma_a, \Gamma)  \right. \right. 
\notag\\
&&
 - \frac{1}{2} \left( \frac{2}{3} \right)^{2/3}
 G(5/3, -n+7/3, \Gamma_a, \Gamma)
 + \left( \frac{3}{2} \right)^{1/3} \Gamma_a
 G(2, -n+3, \Gamma_a, \Gamma)
\notag\\
&&
\left. \left.
 - \frac{1}{2} 
 \left( \frac{2}{3} \right)^{2/3}
 G(1, -n+3, \Gamma_a, \Gamma)  
\right\} \right]
\label{interactioniteration}
\end{eqnarray}
where 
\begin{equation}
 G(n_1, n_2, \Gamma_1, \Gamma_2):= 
 \int_{x_0} dx x^{n_1-1} \exp \left( - \Gamma_1 x \right)
 \int_{x_0}^x dy y^{n_2-1} 
 \exp \left\{ \left( \Gamma_1 - \Gamma_2 \right) y \right\},
\label{doublegammadef}
\end{equation}
where 
\begin{equation}
 x := \frac{\gamma}{A^{1/2}} \frac{2}{3} a^{3/2}, \quad                 
 x_0 := \frac{\gamma}{A^{1/2}} \frac{2}{3}
\end{equation}
The evaluation of the double Gamma function defined by
(\ref{doublegammadef}) is treated in Appendix $B$.
We consider the concrete example defined by
\begin{equation}
 U_{\rm int} = \lambda \phi^2_1 \phi^2_2, \quad
 m_1 = m_2.
\label{interactionmodel}
\end{equation}
In this case, we use the independent variables defined by
\begin{eqnarray}
 && \theta_1 = q_0 \quad 
    \theta_2 = q_0 + q_1
\notag\\
 && I_1 = p_0 - p_1 \quad
    I_2 = p_1,
\end{eqnarray}
where $(q_0, p_0)$ and $(q_1, p_1)$ are called fast and slow
action-angle variables, respectively \cite{Hamazaki2004}.
Because of the resonant relation satisfied by the masses
of the scalar fields, the slow angle variable $q_1$ moves 
much more slowly than the fast angle variable $q_0$.
The averaging over the slow angle variable $q_1$ cannot be
justified, and therefore the slow action-angle variables
$(q_1, p_1)$ can have evolutions.
In the previous paper \cite{Hamazaki2004}, we investigated
the influences of the resonant interaction on the evolution of
the cosmological perturbations before the energy transfer from
the scalar fields to the radiation fluids begins.
According to this study, the slow action-angle variables can 
suffer from the instability near the hyperbolic fixed point
in the phase space of the slow action-angle variables.
Since the initial adiabatic perturbation $\xi$ and 
the initial isocurvature perturbation $\eta_{12}$ are given by
\begin{equation}
 \xi = \frac{1}{3} \frac{\delta p_0}{p_0}
\end{equation}
and
\begin{equation}
 \eta_{12} = 
 \frac{p_1 \delta p_0 - p_0 \delta p_1}{(p_0- p_1) p_1},
\end{equation}
respectively, the instability of the slow action variable $p_1$
has influence on the isocurvature mode.
Therefore from (\ref{Bardeenadiiso})
in the case $\Gamma_1 \neq \Gamma_2$ the instability 
of the action-angle variables survives in the final amplitude
of the Bardeen parameter.
Next we calculate the first order correction term 
(\ref{interactioniteration}) in the present model 
(\ref{interactionmodel}).
The present model has the hyperbolic fixed point at
\begin{equation}
 q_1 = \frac{\pi}{2} (2 k +1), \quad
 2 p_1 = p_0 = c,
\end{equation}
where $k$ is an integer.
At this hyperbolic fixed point, the first order correction 
to the final amplitude of the Bardeen parameter 
(\ref{interactioniteration}) is calculated as
\begin{equation}
 \zeta^{\sharp}_{\rm fin} \supset 
 \frac{1}{12} \frac{1}{A^{1/2}}
 \frac{\nu}{\epsilon}
 c^2 \delta q_1 (1)
\left( 
 \frac{1}{\Gamma^{2/3}_1} -
 \frac{1}{\Gamma^{2/3}_2}
\right)
 \Big/ \sum_a \frac{A_a}{\Gamma^{2/3}_a},
\end{equation}
where 
\begin{equation}
 \nu := \frac{\lambda I_0}{m^3_0 a^3_0}
\end{equation}
and the non-dimensional masses are scaled as $m_1 =m_2 =1$.
In the present model,
as for the first order correction term also, in order that
the slow action-angle variables instability has influence
on the final Bardeen parameter, 
$\Gamma_1 \neq \Gamma_2$ is necessary.

Until now, we evaluate $I_a$ $\sigma_{\alpha}$ by assuming that
$\rho$ is given by (\ref{earlyenergydensity}).
Now we evaluate the contribution to $I_a$ $\sigma_{\alpha}$
from the late stage of reheating when $\rho$ is given by
\begin{equation}
 \rho_{\rm late} =
\left( \frac{3}{2} \right)^{2/3} G \left( 5/3 \right)
 \frac{1}{a^4}
 \frac{A^{1/3}}{\gamma^{2/3}}
 \sum_a \frac{A_a}{\Gamma^{2/3}_a}.
\end{equation}
Such late stage of reheating begins at
\begin{equation}
 a_1 = d 
\left( \frac{3}{2} \right)^{2/3} G \left( 5/3 \right)
 \frac{1}{\gamma^{2/3} A^{2/3}}
 \sum_a \frac{A_a}{\Gamma^{2/3}_a},
\end{equation}
because at this $a_1$, $A/a^3$ is almost equal to $\rho_{\rm late}$.
$d$ is a numerical factor which can be assumed to be larger than unity.
By using $\rho_{\rm late}$, $a_1$, we can evaluate the contribution
to $\rho_r$ from the late stage of reheating as 
\begin{eqnarray}
 \rho_r &\supset&
 \frac{1}{a^4}
\left( \frac{3}{2} \right)^{2/3} G \left( 5/3 \right)
 \sum_a d 
\exp\left\{  
 - d^{3/2} G(5/3)^{3/2}
 \frac{\Gamma_a}{A^{3/2}}
\left(
 \sum_b \frac{A_b}{\Gamma^{2/3}_b}
\right)^{3/2}
     \right\} 
\notag\\
 && \times
 \frac{A_a}{\gamma^{2/3} A^{2/3}}
 \sum_b \frac{A_b}{\Gamma^{2/3}_b},
\end{eqnarray}
whose size is characterized by
\begin{equation}
 r_{l/e} := d 
 \exp\left\{    
 - d^{3/2} G(5/3)^{3/2}
     \right\},
\end{equation}
which is the ratio of the late contribution to the main early
contribution to $\rho_r$.
The value of $r_{l/e}$ is $0.43$, $0.18$ and $0.035$ for 
$d=1,2$ and $3$, respectively.
Therefore we can conclude that the S formula which is derived by 
using (\ref{earlyenergydensity}) is rather good approximation
to the real S formula.

We consider the case where the decay rates $\Gamma_a$ depend 
on the radiation temperature $T$.
In the high temperature limit $T \gg m$ where $m$ is the mass 
scale of the oscillatory scalar fields, the decay rate $\Gamma_a$
depends on the radiation temperature \cite{Yokoyama2004}.
When the decay product is the fermion, $\Gamma_a$ is given by
\begin{equation}
 \Gamma_a = \alpha_a \frac{1}{T}.
\label{fermiondecay}
\end{equation}
When the decay product is the boson, $\Gamma_a$ is given by
\begin{equation}
 \Gamma_a = \beta_a T.
\label{bosondecay}
\end{equation}
According to the paper \cite{Yokoyama2004} the reason is following.
We consider the case where $\rho_r \sim T^4$ is sufficiently high.
In the fermion case, the Pauli exclusion principle inhibits the decay
of $\phi_a$ into fermions since the fermions have already occupied the 
energy levels into which $\phi_a$ would decay.
In the boson case, the induced effect promotes the decay of $\phi_a$
into bosons since the bosons occupy the energy levels into which 
$\phi_a$ decay.
For simplicity, we consider the case where the radiation consists
of one component.
We interpret that the radiation temperature $T$ appearing
in (\ref{fermiondecay})(\ref{bosondecay}) is the temperature
$T_a := \left[ \rho_r (a(\Gamma_a)) \right]^{1/4}$
at the time when the decay process proceeds given by
\begin{equation}
 a (\Gamma_a) := \frac{A^{1/3}}{\gamma^{2/3} \Gamma^{2/3}_a}.
\end{equation}
In the fermion case, by substituting $T_a$ defined above
to (\ref{fermiondecay}) it can be verified that
\begin{equation}
 \sum_a \frac{A_a}{\Gamma^{2/3}_a} =
 \frac{\gamma^{2/9}}{A^{1/9}}
 \left( \sum_a \frac{A_a}{\alpha^{2/5}_a} \right)^{10/9}.
\end{equation}
Therefore we obtain
\begin{eqnarray}
 a^4 \rho_r &\supset& \frac{A^{2/9}}{\gamma^{4/9}}
 \left( \sum_a \frac{A_a}{\alpha^{2/5}_a} \right)^{10/9},\\
 \zeta^{\sharp}_{\rm fin} &\supset& \xi + \frac{5}{36}
 \sum_{ab} \left( 
 \frac{1}{\alpha^{2/5}_a} - \frac{1}{\alpha^{2/5}_b} 
           \right)
 \frac{A_a A_b}{A} \eta_{ab}
 \Big/ \sum_a \frac{A_a}{\alpha^{2/5}_a}.
\end{eqnarray}
In the same way as in the fermion case, in the boson case
we can obtain
\begin{eqnarray}
 a^4 \rho_r &\supset& \frac{A^{2/3}}{\gamma^{4/3}}
 \left( \sum_a \frac{A_a}{\beta^{2}_a} \right)^{2/3},\\
 \zeta^{\sharp}_{\rm fin} &\supset& \xi + \frac{1}{12}
 \sum_{ab} \left( 
 \frac{1}{\beta^{2}_a} - \frac{1}{\beta^{2}_b} 
           \right)
 \frac{A_a A_b}{A} \eta_{ab}
 \Big/ \sum_a \frac{A_a}{\beta^{2}_a}.
\end{eqnarray}
The radiation temperature dependence of the decay rate $\Gamma_a$
affects how the isocurvature modes are transmitted into the 
final amplitude of the Bardeen parameter.

We consider the non-Gaussianity of perturbations. 
For simplicity, we assume that all the perturbations are generated
by only one Gaussian variable, say locally homogeneous perturbed
variable $C(\bm{x}) = C + \delta C(\bm{x})$ where 
$\delta C(\bm{x})$ is spatially dependent Gaussian random variable.
By assuming that $\rho_r (a, C(\bm{x})) \propto C(\bm{x})^{\alpha}$,
from (\ref{bardeengenerator}), the non-linearity parameters 
\cite{Komatsu2001} defined by
\begin{equation}
 \zeta = \zeta_1 + \frac{3}{5} f_{NL} \zeta^2_1 
 + \frac{9}{25} g_{NL} \zeta^3_1 + \cdot \cdot \cdot
\end{equation}
where $\zeta_1$ is the first order perturbation of 
the Bardeen parameter generated by 
one Gaussian random variable, are given by
\begin{equation}
 f_{NL} = - \frac{10}{3 \alpha}, \quad \quad
 g_{NL} = \frac{100}{27 \alpha^2}. 
\end{equation}
In the reheating model with only one scalar field, the final radiation
energy density is given by
\begin{equation}
 a^4 \rho_r \sim \frac{1}{\gamma^{2/3}} \frac{A^{4/3}}{\Gamma^{2/3}}.
\end{equation} 
When the initial action variable $A/m$ is the random Gaussian variable,
by considering $A(\bm{x}) \propto C(\bm{x})^2$ from
(\ref{actiongaussian}), non-linearity parameters are given by
\begin{equation}
 f_{NL} = - \frac{5}{4}, \quad \quad
 g_{NL} = \frac{25}{12}.
\end{equation}
In the modulated reheating scenario, by assuming that the decay rate
$\Gamma(\bm{x})$ is proportional to $\phi(\bm{x})^{\beta}$, the
non-linearity parameters are given by
\begin{equation}
 f_{NL} = \frac{5}{\beta}, \quad \quad
 g_{NL} = \frac{100}{3 \beta^2}.
\end{equation}
By observing the non-Gaussianity, we can determine whether the
nonnegligiable Gaussian random variable lies 
in the action variable $A(\bm{x})/m$
or the decay rate $\Gamma (\bm{x})$.

\section{Application of the LWL formula to the multicomponent curvaton
model}

In this section, we apply the LWL formula to the curvaton scenario where
multiple weakly coupled massive scalar fields called curvatons decay
into multiple radiation fluids some time later after the inflation
has ended.
In this curvaton scenario, the curvaton fields other than the inflaton
fields driving the inflation are responsible for the origin of the 
cosmic structures.
First, we assume that 
all $\Gamma_a := \sum_{\alpha} \Gamma_{\alpha a}$
($a=1,2, \cdot \cdot \cdot ,N_S $) are of the same order of
magnitude.
 
First we consider the limit where in the initial time the curvaton
fields energy $A$ is small compared to the radiation fluids energy
$B$: $A \ll B$.
By substituting the $\rho$ expression (\ref{earlyenergydensity}) to
(\ref{quasisolution1})(\ref{quasisolution2}) and
by expanding it with respect to $A$ around $A=0$, we obtain
\begin{eqnarray}
 m_a I_a &=& A_a 
 \exp \left\{ - \frac{1}{2} \frac{\gamma \Gamma_a}{B^{1/2}}   
 a^{2}
      \right\}
 + O(A^2),
\notag\\
a^4 \rho_r &=& a^4 \sum_{\alpha} \rho_{\alpha} =
 B+ \sqrt{2} G(3/2) 
 \frac{B^{1/4}}{\gamma^{1/2}}
 \sum_a \frac{A_a}{\Gamma^{1/2}_a} + O(A^2).
\label{energyBlarge}
\end{eqnarray}
By taking the exactly homogeneous perturbation, that is $D$ operation
around $A = 0$, we obtain
\begin{equation}
 \zeta^{\sharp}_{\rm fin} \cong 
 \frac{1}{4} \frac{D \rho^{\sharp}}{\rho}
 = \frac{1}{4} \frac{\delta B}{B} + \frac{\sqrt{2}}{4} G(3/2)
 \frac{1}{\gamma^{1/2} B^{3/4}} 
 \sum_a \frac{\delta A_a}{\Gamma^{1/2}_a}.
\label{AsmallSformula}
\end{equation}

Next we consider the case where the energy densities of the curvatons
are large compared with those of radiation fluids when the energy 
transfer from the curvatons to the radiation fluids proceeds.
In this case, the exponent of (\ref{quasisolution1}) is written as
\begin{equation}
 \gamma \Gamma_a \int_1 da \frac{1}{\rho^{1/2} a}
 = \frac{\gamma \Gamma_a}{A^2} 
 \left(
\frac{2}{3} x^{3/2} - 2 B x^{1/2} + \frac{4}{3} B^{3/2}
 \right),
\end{equation}
where 
\begin{equation}
 x := A a + B.
\end{equation}
As for $x(\Gamma_a)$ defined by
\begin{equation}
 \gamma \Gamma_a \int^{\left\{ x(\Gamma_a)-B \right\}/A}_1 
 da \frac{1}{\rho^{1/2} a} =1,
\end{equation}
we obtain
\begin{equation}
 x(\Gamma_a) = x_0 (\Gamma_a)
\left[  
 1+ 2 \frac{B}{x_0 (\Gamma_a)}
 - \frac{4}{3} \left( \frac{B}{x_0 (\Gamma_a)} \right)^{3/2}
 + O \left( \frac{B^2}{x_0 (\Gamma_a)^2} \right)
\right]
\end{equation}
where
\begin{equation}
 x_0 (\Gamma_a) := \left( \frac{3}{2} \right)^{2/3}
 \left( \frac{A^2}{\gamma \Gamma_a} \right)^{2/3}.
\end{equation}
The expansion parameter $B/ x_0(\Gamma_a)$ implies the
ratio of the energy of radiations to the energy of the 
curvaton $\phi_a$ when the energy transfer proceeds.
By using $x(\Gamma_a)$, we approximate the exponential
function by the step function:
\begin{equation}
 \exp\left\{  
 - \gamma \Gamma_a \int_1 da \frac{1}{\rho^{1/2} a}
 \right\}
 \to \theta \left( x(\Gamma_a) -x \right).
\end{equation}
By using this approximation, we obtain the radiation energy 
density in the $a \to \infty$ limit:
\begin{equation}
 \rho_r = \sum_{\alpha} \rho_{\alpha}
 = \frac{1}{a^4} 
    \left[
 B + \frac{2}{5} \left( \frac{3}{2} \right)^{5/3}
 \frac{A^{1/3}}{\gamma^{2/3}} 
 \sum_a \frac{A_a}{\Gamma^{2/3}_a}
 + B + \cdot \cdot \cdot
    \right].
\label{energyAlarge}
\end{equation}
By taking the exactly homogeneous perturbation, we 
obtain the S formula:
\begin{equation}
 \zeta^{\sharp}_{\rm fin} = 
 \frac{1}{4} \frac{\delta A_{\ast}}{A_{\ast}}
 + \frac{1}{2} \frac{\delta B}{A_{\ast}}
 - \frac{B}{2 A^2_{\ast}} \delta A_{\ast}
 + \cdot \cdot \cdot,
\end{equation}
where 
\begin{equation}
 A_{\ast} := \frac{2}{5} \left( \frac{3}{2} \right)^{5/3}
 \frac{A^{1/3}}{\gamma^{2/3}} 
 \sum_a \frac{A_a}{\Gamma^{2/3}_a}.
\end{equation}
We define more physical parametrization by
\begin{eqnarray}
 \delta A_a &=& 3 A_a \xi + 3 A_a \eta + 
 \sum_b \frac{A_a A_b}{A} \eta_{ab},\\
 \delta B_{\alpha} &=& 4 B_{\alpha} \xi + 
 \sum_{\beta} \frac{B_{\alpha} B_{\beta}}{B} \eta_{\alpha \beta}.
\end{eqnarray}
In particular, since
\begin{equation}
 \delta A = 3 A \xi + 3 A \eta, \quad
 \delta B = 4 B \xi,
\end{equation}
$\xi$ implies the adiabatic growing mode and $\eta$ means
the isocurvature mode between the total curvatons and 
the total radiations.
By using this parametrization, the S formula can be rewritten
as 
\begin{eqnarray}
 \zeta^{\sharp}_{\rm fin} &=& \xi +
 \left( 1 - 2 \frac{B}{A_{\ast}} \right) \eta
\notag\\
 && + \left( \frac{1}{8} - \frac{B}{4 A_{\ast}} \right)
 \sum_{ab} \left( 
 \frac{1}{\Gamma^{2/3}_a} - \frac{1}{\Gamma^{2/3}_b} 
           \right)
 \frac{A_a A_b}{A} \eta_{ab} \Big/
 \sum_a \frac{A_a}{\Gamma^{2/3}_a}
 + \cdot \cdot \cdot.
\label{AlargeSformula}
\end{eqnarray} 

In the most simple curvaton scenario of one curvaton field
and one radiation fluid, the empirical S formula was obtained
from the numerical calculation \cite{Gupta2004}:
\begin{eqnarray}
 \zeta_{\rm fin} &=& r(p) \eta,\\
 r(p) &=& 1 - \left( 1+ \frac{0.924}{1.24} p \right)^{-1.24},
\end{eqnarray}
where 
\begin{equation}
 p := \frac{A}{\gamma^{1/2} \Gamma^{1/2} B^{3/4}}.  
\end{equation}
In the limits $p \ll 1$, $p \gg 1$, this empirical S formula
gives 
\begin{alignat}{2}
 \zeta_{\rm fin} &= 0.924 p \eta  & \qquad
 & p \ll 1, \\
 \zeta_{\rm fin} &= \left( 1 - \frac{1.44}{p^{1.24}} \right)
 \eta & \qquad
 & p \gg 1,
\end{alignat}
respectively.
Our analytical results (\ref{AsmallSformula}) (\ref{AlargeSformula})
give
\begin{alignat}{2}
 \zeta^{\sharp}_{\rm fin} &\supset 0.940 p \eta  & \qquad
 & p \ll 1, \\
 \zeta^{\sharp}_{\rm fin} &\supset 
 \left( 1 - \frac{2.54}{p^{1.33}} \right)
 \eta & \qquad
 & p \gg 1,
\end{alignat}
respectively.
In the case $p \ll 1$, the empirical formula and our analytic result
agree with good accuracy.
In the case $p \gg 1$, our analytic result is obtained by rather 
rough treatment approximating the exponential function by the 
step function.
But our analytic S formula agree well with the empirical formula.  
For reference, for $p \gg 1$, according to our method, a more
precice calculation gives 
\begin{eqnarray}
 \zeta^{\sharp}_{\rm fin} &\supset&
 \left[   
 1- \frac{10}{3} \left( \frac{2}{3} \right)^{2/3}
 \frac{1}{p^{4/3}}
 + \frac{10}{3} \frac{1}{p^{2}}
 + \frac{10}{9} \left( \frac{2}{3} \right)^{4/3}
 \frac{1}{p^{8/3}}
 + O \left( \frac{1}{p^{10/3}} \right)
 \right] \eta
\notag\\
 &=&
 \left[   
 1- 2.54 \frac{1}{p^{1.33}}
 + 3.33 \frac{1}{p^{2}}
 + 0.647
 \frac{1}{p^{2.67}}
 + O \left( \frac{1}{p^{3.33}} \right)
 \right] \eta.
\end{eqnarray}
The errors between the above formula and the empirical formula
are $1.3$ percent, $1.1$ percent for $p=10$, $p=5$, respectively.

In section $6$, until now in section $7$, we assumed that all decay
rates $\Gamma_a$ are of the same order of magnitude.
As application example of the formulae (\ref{energyBlarge})
(\ref{energyAlarge}), we consider the reheating where the decay rate
of $\phi_1$ is much larger than the decay rate $\phi_2$;
$\Gamma_1 \gg \Gamma_2$.
Just after the scalar field $\phi_1$ decays, by using the result of 
section $6$, we obtain
\begin{equation}
 a^4 \rho_r \sim \frac{A^{1/3}}{\gamma^{2/3}}
 \frac{A_1}{\Gamma^{2/3}_1}, \quad \quad
 a^3 \rho_S \sim A_2.
\end{equation} 
By regarding $a^4 \rho_r$ in the above as $B$, we can use 
the formulae (\ref{energyBlarge}) (\ref{energyAlarge}) of 
the curvaton scenario.
In the case $A_2 / A_1 \gg (\Gamma_2 / \Gamma_1)^{1/2}$, by 
using (\ref{energyBlarge}), the final radiation energy density
is calculated as
\begin{equation}
 a^4 \rho_r \sim \frac{A^{1/3}}{\gamma^{2/3}}
 \frac{A_1}{\Gamma^{2/3}_1} +
\frac{1}{\gamma^{2/3}}
 \frac{A^{1/12} A^{1/4}_1 A_2}{\Gamma^{1/6}_1 \Gamma^{1/2}_2}.
\end{equation}
In the case $A^{4/3}_2 / A^{1/3} A_1 \gg (\Gamma_2 / \Gamma_1)^{2/3}$,
by using (\ref{energyAlarge}), the final radiation energy density
is calculated as
\begin{equation}
 a^4 \rho_r \sim \frac{A^{1/3}}{\gamma^{2/3}}
 \frac{A_1}{\Gamma^{2/3}_1} +
\frac{1}{\gamma^{2/3}}
 \frac{A^{4/3}_2}{\Gamma^{2/3}_2}.
\end{equation}
By substituting the above two expressions to (\ref{bardeengenerator})
and by expanding with respect to Gaussian random perturbations,
we can obtain the Bardeen parameter of arbitrary order.

\section{Discussion}

In this paper, we constructed the LWL formula expressing the 
long wavelength limit of evolution of cosmological perturbations 
in terms of the corresponding exactly homogeneous perturbations in 
the most general scalar-fluid composite system.
We determined the correction term which corrects the difference
between the long wavelength limit of cosmological perturbations and
the exactly homogeneous perturbations, and 
we showed that the correction term contributes the well known 
adiabatic decaying mode.
It was pointed out that when we extract the long wavelength limits of
evolutions of cosmological perturbations from the exactly homogeneous
variables, the use of the scale factor $a$ as the evolution parameter
is more useful.
The scalar-fluid composite system whose LWL formula is constructed 
in this paper can be 
used to discribe the early stage of the universe such as reheating
after inflation and the curvaton decay in the curvaton scenario, 
when the fluid is assumed 
to be radiation.
In this paper, the LWL formula is applied to the most general case
of reheating and of the curvaton decay containing the multiple
scalar fields and the multiple radiation fluids, and the S formulae
representing the final amplitude of the Bardeen parameter in terms
of the initial adiabatic and entropic perturbations are constructed.
In case where for different $a$, the value of the decay rate $\Gamma_a$ 
is different; that is
$\Gamma_a \neq \Gamma_b$ for $a \neq b$, 
the initial isocurvature modes survive in the final
amplitude of the Bardeen parameter.

We discuss the non-linear generalization of the LWL formalism.
Recently the gradient expansion has been discussed as the 
method for investigating the evolutions of non-linear perturbations
on superhorizon scales \cite{Rigopoulos2003}.
In the lowest order of the gradient expansion, in the zero
curvature slice $\partial a(t, \bm{x}) / \partial \bm{x}^i = 0$,
the evolution equation of the scalar quantity $T$ has exactly 
the same form as that of the exactly homogeneous equation 
of $T$ \cite{Nambu2006}.
But the coefficients of the evolution equation are spatially 
dependent, therefore this evolution equations describes the non-linear
superhorizon scale inhomogeneities.
By using the solution of the exactly homogeneous system with
the scale factor as the evolution parameter $T(a,C)$, the solution
of the locally homogeneous evolution equation is given by 
$T(a,C(\bm{x}))$, where $C(\bm{x})=C + \delta C(\bm{x})$ is spatially 
dependent solution constant.
The spatially dependent perturbation part of $T(a,C(\bm{x}))$
is given by Taylor expanding with respect to 
$C(\bm{x})=C + \delta C(\bm{x})$;
\begin{equation}
 T(a,C(\bm{x})) = T(a,C) + \sum^{\infty}_{k=1} 
 \sum_{a_1} \cdot \cdot \cdot \sum_{a_k} \frac{1}{k!}
 \frac{\partial^k T}
 {\partial C_{a_1} \cdot \cdot \cdot \partial C_{a_k} }
 \delta C_{a_1} (\bm{x}) \cdot \cdot \cdot \delta C_{a_k} (\bm{x}),
\end{equation}
whose first order perturbation part agrees with our linear perturbation
formula with neglecting the adiabatic decaying mode; 
$D T^{\sharp} = (\partial T / \partial C )_a$.
Therefore $T(a,C(\bm{x}))$ is the non-linear generalization of our
linear perturbation variable $D T$. 
When $P_r = \rho_r / 3$ as in the final state of reheating or the 
curvaton decay, $T = \ln {\rho_r} / 4$ is the non-linear generalization
of the Bardeen parameter $\zeta$.
Since in this paper we determine $\rho_r$ in the final state of 
reheating or the curvaton decay with the scale factor as the evolution 
parameter, we can obtain the information of the non-linear evolution
and the non-Gaussianity of perturbations which fluctuate spatially
on superhorizon scales.

Our evolution equations have arbitrary functions $S$ which describe
the energy transfer between scalar fields and perfect fluids.
The source functions $S$ can be determined from the microscopic 
dynamics between the coherently oscillating scalar fields and radiation,
concretely speaking, by path integrating out the fields constituting the
radiation and interacting with the coherently oscillating scalar fields
in the effective in-in action \cite{Yokoyama2004}.
When the scalar fields oscillate coherently, the source term such as
$S = \Gamma \dot{\phi}$ is important in order that the energy is
transferred from the scalar fields into radiation effectively.
As shown in the paper \cite{Yokoyama2004}, $\Gamma$ is given as the 
function $\Gamma = \Gamma (\phi, \dot{\phi}, \rho_r)$ where $\rho_r$ 
is the energy density of radiation.
As seen in the section $6$,
when $\Gamma$ is a function of the scalar 
quantities which are closely related to reheating process,
the functional form of $\Gamma$ affects how the initial isocurvature
components are converted into the adiabatic component such as 
the final amplitude of the Bardeen parameter, but it does not
affect the evolution of the initial adiabatic growing mode.
On the other hand,
the scenario where the energy transfer is controlled by the 
scalar quantities not related to reheating is considered as the 
modulated reheating scenario \cite{Sasaki1991} \cite{Dvali2004}.
As such scalar quantities, we can choose flat direction scalar field
which does not govern the energy of the universe but fluctuates
of the order of the Hubble parameter during the inflationary expansion,
or the scalar field written in terms of the slow action variable which
suffers from the hyperbolic instability due to the resonance of the 
masses of the scalar fields during the oscillatory stage 
\cite{Hamazaki2004}.
As seen from the formula (\ref{modulatereheating})
the fluctuations of such modulating scalar quantities are
imprinted on the $\rho_r$ fluctuation.

We consider the system where multiple oscillatory scalar fields and
multiple radiation fluids interact.
As for the system without radiation, its evolution of cosmological 
perturbations have been investigated in detail \cite{Hamazaki2002},
\cite{Hamazaki2004}. 
The system of multiple scalar fields only can be written in terms
of Hamiltonian form.
Any Hamiltonian system obeys the Liouville theorem, that is, volume
occupied by group of orbits are invariant.
Therefore according to the LWL formula, in the stable case 
the perturbations do not grow and in the unstable case the same numbers
of growing modes and decaying modes appear because of the squeezing of
the phase space volume.
The former case occurs in the case where masses of scalar fields are
incommensurable and near the elliptic fixed points in the case where
masses of scalar fields are commensurable.
The latter case occurs near the hyperbolic fixed points in the case 
where masses of scalar fields are commensurable.
Then we include dissipative interaction with radiation.
In this case our system is not a Hamiltonian system and it does not
obey the Liouville theorem.
In this dissipative system, in addition to two possibilities mentioned
above we can expect the third possibility where group of orbits are 
attracted into the attracting set.
The final state where all the energy of the scalar fields is transferred
into that of radiation fluids is the attracting equilibrium.
Around the attracting set, the adjacent orbits come nearer and nearer,
therefore the LWL formula tells us that all the perturbation modes
are stable, that is, converge into some constants or decay.
It is useful to investigate how the behavior of cosmological
perturbations around the hyperbolic fixed points is changed due to
the dissipative interaction with radiation under the spirit of the 
LWL formula.
In this line of researches, we will have
new explanation of the backreaction which suppresses the instability
due to the resonance.
In the future publication, we will return to this problem.

\section*{Acknowledgments}

The author would like to thank Professors H. Kodama, S. Mukohyama,
T. Nakamura, M. Sasaki, K. Sato, N. Sugiyama, T. Tanaka, A. Taruya, 
J. Yokoyama for continuous encouragements.
He would like to thank Professor V.I. Arnold for writing his
excellent textbook and/or review \cite{Arnold}, from which
he learned a lot about the dynamical system.

\appendix


\catcode`\@=11

\@addtoreset{equation}{section}   
\def\theequation{\Alph{section}.\arabic{equation}}


\section{Proofs of the propositions in \S $5$}

\subsection{Technical Lemmas}

\paragraph{Technical Lemma $1$}

\textit{Under the assumption (\ref{assumptiontwo}), for
$|f| \le 1$,}
\begin{equation}
 \Big| \frac{\partial f}{\partial I} \Big| \le a, \quad
 \Big| \frac{\partial f}{\partial a} \Big| \le \frac{1}{a},
\end{equation}
\textit{and for $|f| \le |I|$,}
\begin{equation}
 \Big| \frac{\partial f}{\partial I} \Big| \le 1, \quad
 \Big| \frac{\partial f}{\partial a} \Big| \le \frac{|I|}{a},
\end{equation}

\paragraph{Technical Lemma $2$}

\textit{For the general physical quantity $A(I, \sigma, \theta)$
with Fourier decomposition as}
\begin{equation}
 A = A_0 (I, \sigma) + 
 \sum_{\bm{k} \ne \bm{0}} A_{\bm{k}} (I, \sigma)
 \exp{ (i \bm{k} \cdot \bm{\theta}) },
\end{equation}
\textit{the solution to the first order partial differential
equation}
\begin{equation}
 \omega_a \frac{\partial}{\partial \theta_a} S =
 A- <A>,
\end{equation}
\textit{is given by}
\begin{equation}
 S = \{ A \},
\end{equation}
\textit{where}
\begin{equation}
 \{ A \} := \sum_{\bm{k} \ne \bm{0}} 
 \frac{A_{\bm{k}}}{i (\bm{k} \cdot \bm{\omega})}
 \exp{ (i \bm{k} \cdot \bm{\theta}) }. 
\end{equation}

\subsection{Proof of Proposition $1$}

In order to make the notation simple, we omit the superscript $(k)$
and replace $(k+1)$ with $(1)$.
By substituting the transformation laws of $I$ $\sigma$ $\theta$
to the evolution equations of $I_a$ $\sigma_{\alpha}$ $\theta_a$,
we obtain
\begin{eqnarray}
 && <F_a> + F_a (I, \sigma, \theta, a)
 - F_a (I^{(1)}, \sigma^{(1)}, \theta^{(1)}, a)
 - a \frac{\partial u_a}{\partial a}
\notag\\
 &&= F^{(1)}_a 
+\frac{\partial u_a}{\partial I^{(1)}_b} F^{(1)}_b
+\frac{\partial u_a}{\partial \sigma^{(1)}_{\beta}} F^{(1)}_{\beta}
+\frac{\partial u_a}{\partial \theta^{(1)}_b} G^{(1)}_b,
\label{Itrans} 
\end{eqnarray}
\begin{eqnarray}
 && <F_{\alpha}> + F_{\alpha} (I, \sigma, \theta, a)
 - F_{\alpha} (I^{(1)}, \sigma^{(1)}, \theta^{(1)}, a)
 - a \frac{\partial u_{\alpha}}{\partial a}
\notag\\
 &&= F^{(1)}_{\alpha} 
+\frac{\partial u_{\alpha}}{\partial I^{(1)}_b} F^{(1)}_b
+\frac{\partial u_{\alpha}}{\partial \sigma^{(1)}_{\beta}} F^{(1)}_{\beta}
+\frac{\partial u_{\alpha}}{\partial \theta^{(1)}_b} G^{(1)}_b,
\label{sigmatrans} 
\end{eqnarray}
and
\begin{eqnarray}
&&
 \frac{1}{\epsilon} \omega_a (I, \sigma, a)
-\frac{1}{\epsilon} \omega_a (I^{(1)}, \sigma^{(1)}, a)
-\frac{1}{\epsilon} \frac{\partial \omega_a}{\partial I^{(1)}_b} u_b
-\frac{1}{\epsilon} 
 \frac{\partial \omega_a}{\partial \sigma^{(1)}_{\beta}} u_{\beta}
\notag\\
 && + <G_a> + G_a (I, \sigma, \theta, a)
 - G_a (I^{(1)}, \sigma^{(1)}, \theta^{(1)}, a)
 - a \frac{\partial v_a}{\partial a}
\notag\\
 &&= G^{(1)}_a 
+\frac{\partial v_a}{\partial I^{(1)}_b} F^{(1)}_b
+\frac{\partial v_a}{\partial \sigma^{(1)}_{\beta}} F^{(1)}_{\beta}
+\frac{\partial v_a}{\partial \theta^{(1)}_b} G^{(1)}_b,
\label{thetatrans} 
\end{eqnarray}
when we choose $u_a$ $u_{\alpha}$ $v_a$ as
\begin{eqnarray}
 u_a &=& \epsilon \{ F_a \} \sim \epsilon^{k+1} |I|,\\
 u_{\alpha} &=& \epsilon \{ F_{\alpha} \} \sim \epsilon^{k+1} a |I|,\\
 v_a &=&
 \left\{ 
  \frac{\partial \omega_a}{\partial I^{(1)}_b} u_b
+ \frac{\partial \omega_a}{\partial \sigma^{(1)}_{\beta}} u_{\beta}
+ \epsilon G_a \right\} \sim \epsilon^{k+1}.
\end{eqnarray}
As for $\Delta F_a$ $\Delta F_{\alpha}$ $\Delta G_a$ defined by
\begin{eqnarray}
 \Delta F_a  &:=& F^{(1)}_a - <F_a>,\\
 \Delta F_{\alpha} &:=& F^{(1)}_{\alpha} - <F_{\alpha}>,\\
 \Delta G_a  &:=& G^{(1)}_a - <G_a>,
\end{eqnarray}
applying the mean value theorem to (\ref{Itrans}) (\ref{sigmatrans}) 
(\ref{thetatrans}) gives
\begin{eqnarray}
 && \Delta F_a + \epsilon^{k+1} \Delta F_b 
+ \epsilon^{k+1} |I| \Delta F_{\beta}
+ \epsilon^{k+1} |I| \Delta G_b
= \epsilon^{k+1} a^2 |I|,\\
 && \Delta F_{\alpha} + \epsilon^{k+1} a \Delta F_b 
+ \epsilon^{k+1} a |I| \Delta F_{\beta}
+ \epsilon^{k+1} a |I| \Delta G_b
= \epsilon^{k+1} a^3 |I|,\\
 && \Delta G_a + \epsilon^{k+1} a \Delta F_b 
+ \epsilon^{k+1} \Delta F_{\beta}
+ \epsilon^{k+1} \Delta G_b
= \epsilon^{k+1} a^2,
\end{eqnarray}
where all coefficients of order unity are omitted.
By solving the above three equations, we obtain
\begin{eqnarray}
 \Delta F_a &=& \epsilon^{k+1} a^2 |I|,\\
 \Delta F_{\alpha} &=& \epsilon^{k+1} a^3 |I|,\\
 \Delta G_a &=& \epsilon^{k+1} a^2,
\end{eqnarray}
$\Delta F_a$, $\Delta F_{\alpha}$, $\Delta G_a$ are decomposed
into the angle variables independent parts
$\Delta <F_a>$, $\Delta <F_{\alpha}>$, $\Delta <G_a>$
and the angle variables dependent parts
$\tilde{F}^{(1)}_a$ $\tilde{F}^{(1)}_{\alpha}$ $\tilde{G}^{(1)}_a$.

\subsection{Lemmas and the preparatory propositions}

\paragraph{Lemma $1$}

\textit{The solution to the differential equation}
\begin{equation}
 a \frac{d}{d a} A = - \lambda a^2 A + E(- a^2) B(a)
\end{equation}
\textit{where $\lambda$ is a positive constant, is bounded as}
\begin{equation}
 |A(a)| \le \exp{ \left( - \frac{\lambda}{2} a^2 \right) } |A(1)|
 + E(- a^2) \| B \| (a)
\end{equation}

\paragraph{Lemma $2$}

\textit{When $B$ satisfies}
\begin{equation}
 \Big| \frac{d}{d a} B \Big| \le E(-a^2) |B| 
 + E(-a^2) \{ C + \| B \|(a) \} 
\end{equation}
\textit{where $C$ is a positive constant, for an arbitrary $a \ge 1$}
\begin{equation}
 \| B \|(a) \le \frac{a_1}{a_1 -1}
 \left( C + \| B \|(a_1) \right),
\label{upperB}
\end{equation}
\textit{where $a_1$ is a constant satisfying $a_1 > 1$.
For example, by putting $a_1 = 2$ we obtain}
\begin{equation}
 \| B \|(a) \le C + \| B \|(2).
\end{equation}

\paragraph{Proof}

By solving the differential equation, we obtain
\begin{equation}
 |B(a)| \le |B(1)| + \int_1 d a E(-a^2) \{ C + \| B \|(a) \}.
\end{equation}
Since for $a$ satisfying $1 \le a \le a_1$,
\begin{equation}
 |B(a)| \le C + \| B \|(a_1),
\end{equation}
and for $a \ge a_1$
\begin{equation}
 |B(a)| \le C + \| B \|(a_1) + E(-a^2_1) \| B \|(a), 
\end{equation}
then for an arbitrary $a \ge 1$
\begin{equation}
 \| B(a) \| \le C + \| B \|(a_1) + E(-a^2_1) \| B \|(a), 
\end{equation}
whose right hand side is an increasing function of $a$.
Since $E(-a^2_1) \le 1 / a_1$, we obtain (\ref{upperB}).

\paragraph{Lemma $3$}

\textit{When $\delta I$, $\delta \sigma$ satisfy}
\begin{eqnarray}
 \Big| \frac{d}{da} \delta I \Big| &\le& a |\delta I|
 + |\delta \sigma| +|A|,\\
 \Big| \frac{d}{da} \delta \sigma \Big| &\le& a^2 |\delta I|
 + a |\delta \sigma| +|B|,
\end{eqnarray}
\textit{the following inequality hold:}
\begin{equation}
 \| \delta \sigma \| (2) \le 
 |\delta I(1)|+|\delta \sigma (1)|
 + \| A \|(2)  + \| B \|(2).
\end{equation}

\paragraph{Proof}

We consider the differential equation as for $a |\delta I| 
+ |\delta \sigma|$:
\begin{equation}
 \frac{d}{da} \left( a |\delta I| + |\delta \sigma| \right) 
\le a \left( a |\delta I| + |\delta \sigma| \right) 
+ a |A| + |B|.
\end{equation}
Then we obtain
\begin{equation}
 \| \delta \sigma \| (a) \le \exp{\left( \frac{1}{2} a^2 \right)}
 \{ |\delta I(1)|+|\delta \sigma (1)|
 +a^2 \| A \|(a) + a \| B \|(a) \}.
\end{equation}
We put $a=2$.

\paragraph{Proposition $Ap1$}

\textit{For the $m$-th order system, for the background quantities, 
the following inequalities hold:}
\begin{equation}
 |I| \le \exp{(- a^2)}, \quad
 |\sigma| \le 1.
\end{equation}

\paragraph{Proof}

We solve the evolution equations given by
\begin{eqnarray}
 \Big| a \frac{d}{da} I \Big| &\le& - a^2 |I|
 + a^2 \epsilon^m |I|,\\
 \Big| a \frac{d}{da} \sigma \Big| &\le& a^3 |I|
 + a^3 \epsilon^m |I|.
\end{eqnarray}

\paragraph{Proposition $Ap2$}

\textit{Let $m$ be an integer larger than or equal to $2$.
For the $m$-th order system, for the perturbation quantities,
the following inequalities hold:}
\begin{eqnarray}
 |\delta I (a)| &\le& E(-a^2) \delta A_m (1),\\
 \| \delta \sigma \| (a) &\le& \delta A_m (1),\\
 |\delta \theta (a)| &\le& \exp{(a^2 \epsilon^m)} 
\left\{ |\delta \theta (1)| + \frac{a^2}{\epsilon} \delta A_m (1)    
\right\},
\end{eqnarray}
\textit{where $\delta A_m (1)$ is defined by (\ref{deltaAm}).}

\paragraph{Proof}

The perturbation variables satisfy the evolution equations as
\begin{eqnarray}
 \frac{d}{da} \delta I &=& - a \delta I
 + a |I| \delta \sigma +a \epsilon^m |I| \delta \theta,
\label{deltaIpert}\\
 \frac{d}{da} \delta \sigma &=& a^2 \delta I
 + a^2 |I| \delta \sigma +a^2 \epsilon^m |I| \delta \theta,
\label{deltasigmapert}\\
 \frac{d}{da} \delta \theta &=& \frac{a^2}{\epsilon} \delta I
 + \frac{a}{\epsilon} \delta \sigma +a \epsilon^m \delta \theta,
\label{deltathetapert}
\end{eqnarray}
where $|I|$ means a function bounded by $M |I|$ for a positive
constant $M$.
It is important to notice the coefficient of $\delta I$ in
(\ref{deltaIpert}) is negative.
Although in (\ref{deltaIpert}), terms such as $a^2 |I| \delta I$
is contained, such terms can be neglected because
\begin{equation}
 \int_1 da \> a^2 |I| \le 1.
\end{equation}
By substituting the estimation obtained from (\ref{deltathetapert})
\begin{equation}
 |\delta \theta| \le \exp{(a^2 \epsilon^m)}
 \left\{ |\delta \theta (1)|
 + \frac{a^3}{\epsilon} \| \delta I \|(a)  
 + \frac{a^2}{\epsilon} \| \delta \sigma \|(a)  
 \right\},
\end{equation}
into (\ref{deltaIpert}), by applying Lemma $1$, we obtain
\begin{equation}
 |\delta I (a)| \le E(-a^2) 
 \left\{ 
|\delta I(1)| + \epsilon^m |\delta \theta (1)|
+ \| \delta \sigma \|(a)
 \right\}.
\end{equation}
By using the above inequalities in (\ref{deltasigmapert})
and applying Lemma $2$, we obtain
\begin{equation}
 \| \delta \sigma \|(a) \le  
|\delta I(1)| + \epsilon^m |\delta \theta (1)|
+ \| \delta \sigma \|(2).
\end{equation}
On the other hand, by applying Lemma $3$, we obtain
\begin{equation}
 \| \delta \sigma \|(2) \le \delta A_m (1).
\end{equation}
Then we can prove the results.

\subsection{Proof of Proposition $2A$}

The evolution equations of $\Delta A$ where $A=(I, \sigma, \theta)$
are given by
\begin{eqnarray}
 \frac{d}{da} \Delta I &=& - a \Delta I
 + a |I| \Delta \sigma +a \epsilon^m |I|,\\
 \frac{d}{da} \Delta \sigma &=& a^2 \Delta I
 + a^2 |I| \Delta \sigma +a^2 \epsilon^m |I|,\\
 \frac{d}{da} \Delta \theta &=& \frac{a^2}{\epsilon} \Delta I
 + \frac{a}{\epsilon} \Delta \sigma +a \epsilon^m.
\end{eqnarray}
In the same way as the proof of the Proposition $Ap2$, 
we obtain 
\begin{eqnarray}
 |\Delta I| &\le& E(-a^2) 
 \left(
|\Delta I(1)| +|\Delta \sigma (1)| +\epsilon^m
 \right),\\ 
 \| \Delta \sigma \|(a) &\le&  
|\Delta I(1)| +|\Delta \sigma (1)| +\epsilon^m,\\ 
 |\Delta \theta| &\le& |\Delta \theta(1)|
+ \frac{a^2}{\epsilon}
 \left(
|\Delta I(1)| +|\Delta \sigma (1)| +\epsilon^m
 \right),
\end{eqnarray}
In the above estimations, we put $\Delta A(1)=0$.

Next we consider the evolutions of $\Delta \delta A$
where $A=(I, \sigma, \theta)$.
We take differences between the upper equations and the 
lower equations.
As for $\Delta \delta I$
\begin{eqnarray}
 \frac{d}{da} \delta I &=& - a \delta I
 + a |I| \delta \sigma +a \epsilon^m 
\left( \delta I + |I| \delta \sigma +|I| \delta \theta
\right),\\
 \frac{d}{da} \delta I_{\rm tr} &=& - a \delta I_{\rm tr}
 + a |I_{\rm tr}| \delta \sigma_{\rm tr},
\end{eqnarray}
as for $\Delta \delta \sigma$
\begin{eqnarray}
 \frac{d}{da} \delta \sigma &=& a^2 \delta I
 + a^2 |I| \delta \sigma +a^2 \epsilon^m 
\left( \delta I + |I| \delta \sigma +|I| \delta \theta
\right),\\
 \frac{d}{da} \delta \sigma_{\rm tr} &=& a^2 \delta I_{\rm tr}
 + a^2 |I_{\rm tr}| \delta \sigma_{\rm tr},
\end{eqnarray}
and as for $\Delta \delta \theta$
\begin{eqnarray}
 \frac{d}{da} \delta \theta &=& \frac{a^2}{\epsilon} \delta I
 + \frac{a}{\epsilon} \delta \sigma +a \epsilon^m 
\left( a \delta I + \delta \sigma + \delta \theta
\right),\\
 \frac{d}{da} \delta \theta_{\rm tr} &=& 
   \frac{a^2}{\epsilon} \delta I_{\rm tr}
 + \frac{a}{\epsilon} \delta \sigma_{\rm tr}.
\end{eqnarray}
By taking into account the fact that in the first two terms
on the right hand sides the coefficients depending
on $I$, $\sigma$, not on $\theta$ are multiplied, and
using the estimations of $\delta A$, $\Delta A$ 
where $A=(I, \sigma)$, we obtain
\begin{eqnarray}
 \delta I - \delta I_{\rm tr} &=&
 \Delta \delta I + \delta I (a \Delta I + \Delta \sigma)
\notag\\
 &=&
 \Delta \delta I + E(-a^2) \delta A_m (1) \epsilon^m,\\
 I \delta \sigma  - I_{\rm tr} \delta \sigma_{\rm tr} &=&
 I_{\rm tr} \Delta \delta \sigma + \delta \sigma 
 (\Delta I + |I_{\rm tr}| \Delta \sigma)
\notag\\
 &=&
 E(-a^2) \Delta \delta \sigma + E(-a^2) \delta A_m (1) \epsilon^m,\\
 \delta \sigma - \delta \sigma_{\rm tr} &=&
 \Delta \delta \sigma + \delta \sigma (a \Delta I + \Delta \sigma)
\notag\\
 &=&
 \Delta \delta \sigma + \delta A_m (1) \epsilon^m.
\end{eqnarray}
By using estimations of $\delta A$ where $A=(I, \sigma, \theta)$,
we obtain
\begin{eqnarray}
 && \delta I + |I| \delta \sigma + |I| \delta \theta
= \frac{1}{\epsilon} E(-a^2) \delta A_1 (1),\\
 && a \delta I + \delta \sigma + \delta \theta
= \exp{(a^2 \epsilon^m)}
\left( |\delta \theta (1)| 
  + \frac{a^2}{\epsilon} \delta A_m (1)     
\right).
\end{eqnarray}
Therefore we get
\begin{eqnarray}
 \frac{d}{da} \Delta \delta I &=& - a \Delta \delta I
 + E(-a^2) \Delta \delta \sigma 
 + \epsilon^{m-1} E(-a^2) \delta A_1 (1),\\
 \frac{d}{da} \Delta \delta \sigma &=& a^2 \Delta \delta I
 + E(-a^2) \Delta \delta \sigma 
 + \epsilon^{m-1} E(-a^2) \delta A_1 (1),\\
 \frac{d}{da} \Delta \delta \theta &=& 
 \frac{a^2}{\epsilon} \Delta \delta I
 + \frac{a}{\epsilon} \Delta \delta \sigma \notag\\
&& + a \epsilon^{m-1} \exp{(a^2 \epsilon^m)}
\left( \epsilon |\delta \theta(1)| + a^2 \delta A_m (1)    
\right).
\end{eqnarray}
In the same way as the proof of the Proposition $Ap2$, 
we obtain 
\begin{eqnarray}
 |\Delta \delta I| &\le& E(-a^2) 
 \left(
|\Delta \delta I(1)| +|\Delta \delta \sigma (1)| 
+ \epsilon^{m-1} \delta A_1 (1)
 \right),\\ 
 \| \Delta \delta \sigma \|(a) &\le&  
|\Delta \delta I(1)| +|\Delta \delta \sigma (1)| 
+\epsilon^{m-1} \delta A_1 (1),\\ 
 |\Delta \delta \theta| &\le& |\Delta \delta \theta(1)|
+ \frac{a^2}{\epsilon}
 \left(
|\Delta \delta I(1)| +|\Delta \delta \sigma (1)| 
+ \epsilon^{m-1} \delta A_1 (1)
 \right) \notag\\
&& + \epsilon^{m-1} \exp{(a^2 \epsilon^{m})}
\left(
a^2 \epsilon |\delta \theta (1)| + a^4 \delta A_m (1)
\right).
\end{eqnarray}
In the above estimations, we put $\Delta \delta A(1)=0$
where $A=(I, \sigma, \theta)$.
Then we complete the proof.

\subsection{Proof of Proposition $2B$}

From Proposition $1$, we obtain
\begin{eqnarray}
 I^{(0)} &=& I^{(m)} + \epsilon |I^{(m)}|,\\
 \sigma^{(0)} &=& \sigma^{(m)} + \epsilon a |I^{(m)}|,\\
 \theta^{(0)} &=& \theta^{(m)} + \epsilon,
\end{eqnarray}
where $|I^{(m)}|$ means the function of $A^{(m)}$ where
$A=(I, \sigma, \theta)$ bounded by $M |I^{(m)}|$ for a positive
constant $M$.
As for $\Delta A$ where $A=(I, \sigma, \theta)$, we obtain
\begin{eqnarray}
 |\Delta I^{(0)}- \Delta I^{(m)}| &\le&
\epsilon \left( 
 |\Delta I^{(m)}|+ |I^{(m)}| |\Delta \sigma^{(m)}|
 + |I^{(m)}| |\Delta \theta^{(m)}|
\right)\notag\\
&\le& \epsilon^m E(-a^2),\\
 |\Delta \sigma^{(0)}- \Delta \sigma^{(m)}| &\le&
\epsilon a \left( 
 |\Delta I^{(m)}|+ |I^{(m)}| |\Delta \sigma^{(m)}|
 + |I^{(m)}| |\Delta \theta^{(m)}|
\right)\notag\\   
&\le& \epsilon^m E(-a^2),\\
 |\Delta \theta^{(0)}- \Delta \theta^{(m)}| &\le&
\epsilon \left( 
 a |\Delta I^{(m)}|+ |\Delta \sigma^{(m)}|+ |\Delta \theta^{(m)}|
\right)\notag\\
&\le& \epsilon^m a^2.
\end{eqnarray}
By using the estimations of Proposition $2A$ and the above 
evaluations, we obtain the results of the former part.

Next we consider $\Delta \delta A$ where $A=(I, \sigma, \theta)$.
By taking the variations of the transformation laws, we obtain
\begin{eqnarray}
 \delta I^{(0)}- \delta I^{(m)} &=&
\epsilon \left( 
 \delta I^{(m)}+ |I^{(m)}| \delta \sigma^{(m)}
 + |I^{(m)}| \delta \theta^{(m)}
\right),\\
 \delta \sigma^{(0)}- \delta \sigma^{(m)} &=&
\epsilon a \left( 
 \delta I^{(m)}+ |I^{(m)}| \delta \sigma^{(m)}
 + |I^{(m)}| \delta \theta^{(m)}
\right),\\
 \delta \theta^{(0)}- \delta \theta^{(m)} &=&
\epsilon \left( 
 a \delta I^{(m)}+ \delta \sigma^{(m)}
 + \delta \theta^{(m)}
\right),
\end{eqnarray}
where the coefficients are the functions of $A^{(m)}$
where $A=(I, \sigma, \theta)$.
We take the differences of the transformation laws of the 
exact variables $\delta A$ and those of the truncated variables
$\delta A_{\rm tr}$. 
By using
\begin{eqnarray}
 \delta I^{(m)} - \delta I^{(m)}_{\rm tr} &=&
 \Delta \delta I^{(m)} + \delta I^{(m)} 
 (a \Delta I^{(m)} + \Delta \sigma^{(m)} 
 + \Delta \theta^{(m)}),\\
 I \delta \sigma^{(m)}  - I_{\rm tr} \delta \sigma^{(m)}_{\rm tr} 
 &=&
 I_{\rm tr} \Delta \delta \sigma^{(m)} + \delta \sigma^{(m)} 
 (\Delta I^{(m)} + I_{\rm tr} \Delta \sigma^{(m)}
+I_{\rm tr} \Delta \theta^{(m)}),\\
 \delta \sigma^{(m)} - \delta \sigma^{(m)}_{\rm tr} &=&
 \Delta \delta \sigma^{(m)} + \delta \sigma^{(m)} 
 (a \Delta I^{(m)} + \Delta \sigma^{(m)} + \Delta \theta^{(m)})
,\\
 \delta \theta^{(m)} - \delta \theta^{(m)}_{\rm tr} &=&
 \Delta \delta \theta^{(m)} + \delta \theta^{(m)} 
 (a \Delta I^{(m)} + \Delta \sigma^{(m)} + \Delta \theta^{(m)})
,\\
 I \delta \theta^{(m)}  - I_{\rm tr} \delta \theta^{(m)}_{\rm tr} 
 &=&
 I_{\rm tr} \Delta \delta \theta^{(m)} + \delta \theta^{(m)} 
 (\Delta I^{(m)} + I_{\rm tr} \Delta \sigma^{(m)}
+I_{\rm tr} \Delta \theta^{(m)}),
\end{eqnarray}
and by using the estimations of $\Delta A^{(m)}$ and
$\delta A^{(m)}$ where $A=(I, \sigma, \theta)$, we obtain
\begin{eqnarray}
 \Delta \delta I^{(0)} - \Delta \delta I^{(m)} &=& 
 \epsilon^{m-1} E(-a^2) \delta A_1 (1),\\
 \Delta \delta \sigma^{(0)} - \Delta \delta \sigma^{(m)} &=& 
 \epsilon^{m-1} E(-a^2) \delta A_1 (1),\\
 \Delta \delta \theta^{(0)} - \Delta \delta \theta^{(m)} &=&
 \exp{(a^2 \epsilon^{m})} \epsilon^{m-1}
\left( a^2 \epsilon |\delta \theta (1)| + a^4 \delta A_m (1)
\right).
\end{eqnarray}
By using the estimations of Proposition $2A$ and the above evaluations,
we obtain the results of the latter part.
We complete the proof.

\subsection{Proof of Proposition $3$}

As for $\Delta A$ where $A=(I, \sigma, \theta)$, we can obtain
the results by putting $m=1$ in the proof of Proposition $2A$.

Next we consider the evolutions of $\Delta \delta A$
where $A=(I, \sigma, \theta)$.
We take differences between the upper equations and the 
lower equations.
As for $\Delta \delta I$
\begin{eqnarray}
 \frac{d}{da} \delta I_{\rm tr} &=& - a \delta I_{\rm tr}
 + a |I_{\rm tr}| \delta \sigma_{\rm tr} +a \epsilon 
\left( \delta I_{\rm tr} + |I_{\rm tr}| \delta \sigma_{\rm tr} 
\right),\\
 \frac{d}{da} \delta I_{\rm n} &=& - a \delta I_{\rm n}
 + a |I_{\rm n}| \delta \sigma_{\rm n},
\end{eqnarray}
as for $\Delta \delta \sigma$
\begin{eqnarray}
 \frac{d}{da} \delta \sigma_{\rm tr} &=& a^2 \delta I_{\rm tr}
 + a^2 |I_{\rm tr}| \delta \sigma_{\rm tr} +a^2 \epsilon 
\left( \delta I_{\rm tr} + |I_{\rm tr}| \delta \sigma_{\rm tr} 
\right),\\
 \frac{d}{da} \delta \sigma_{\rm n} &=& a^2 \delta I_{\rm n}
 + a^2 |I_{\rm n}| \delta \sigma_{\rm n},
\end{eqnarray}
and as for $\Delta \delta \theta$
\begin{eqnarray}
 \frac{d}{da} \delta \theta_{\rm tr} &=& 
   \frac{a^2}{\epsilon} \delta I_{\rm tr}
 + \frac{a}{\epsilon} \delta \sigma_{\rm tr} 
 +a \epsilon 
\left( a \delta I_{\rm tr} + \delta \sigma_{\rm tr}
\right),\\
 \frac{d}{da} \delta \theta_{\rm n} &=& 
   \frac{a^2}{\epsilon} \delta I_{\rm n}
 + \frac{a}{\epsilon} \delta \sigma_{\rm n}.
\end{eqnarray}
In the above equations, the subscript ${\rm tr}$ implies
that in the present system, the angle variables dependent 
parts which have been made sufficiently small by the 
transformations defined in the proof of Proposition $1$
have already been truncated, and the subscript ${\rm n}$
means the further neglection of $\epsilon$-order corrections
produced by such transformations.  
We obtain the evolution equations:
\begin{eqnarray}
 \frac{d}{da} \Delta \delta I &=& - a \Delta \delta I
 + E(-a^2) \left( \Delta \delta \sigma 
 + \epsilon \delta B (1) \right),\\
 \frac{d}{da} \Delta \delta \sigma &=& a^2 \Delta \delta I
 + E(-a^2) \left( \Delta \delta \sigma 
 + \epsilon \delta B (1) \right),\\
 \frac{d}{da} \Delta \delta \theta &=& 
 \frac{a^2}{\epsilon} \Delta \delta I
 + \frac{a}{\epsilon} \Delta \delta \sigma 
 + a \delta B(1).
\end{eqnarray}
In the same way as the proof of the Proposition $Ap2$, 
we obtain 
\begin{eqnarray}
 |\Delta \delta I| &\le& E(-a^2) 
 \left(
|\Delta \delta I(1)| +|\Delta \delta \sigma (1)| 
+ \epsilon \delta B (1)
 \right),\\ 
 \| \Delta \delta \sigma \|(a) &\le&  
|\Delta \delta I(1)| +|\Delta \delta \sigma (1)| 
+\epsilon \delta B (1),\\ 
 |\Delta \delta \theta| &\le& |\Delta \delta \theta(1)|
+ \frac{a^2}{\epsilon}
 \left(
|\Delta \delta I(1)| +|\Delta \delta \sigma (1)| 
+ \epsilon \delta B (1)
 \right).
\end{eqnarray}
By putting $\Delta \delta A(1)=0$ where $A=(I, \sigma, \theta)$
in the above inequalities, we obtain the results of the latter
part. 
We complete the proof.

\section{Evaluation of the gamma-like function}

In the present paper, we often have to evaluate the integrals 
defined by
\begin{equation}
 G(t, \Gamma):= \int^{\infty}_{x_0} dx x^{t-1} e^{- \Gamma x},
\end{equation}
where $x_0$ is defined by
\begin{equation}
 x_0 := \frac{2}{3} \frac{\gamma}{A^{1/2}},
\end{equation}
where $\gamma$ is assumed to be sufficiently small.
By expanding with respect to the small parameter $x_0$
by the partial integration, we obtain the evaluations as follows.
For $t > 0$,
\begin{equation}
 G(t, \Gamma) = \frac{1}{\Gamma^t} G(t) =O(1),
\end{equation}
for $t=0$,
\begin{equation}
 G(0, \Gamma) = - \ln{x_0} + O(1),
\end{equation}
for $t < 0$ and $t$ is an integer,
\begin{equation}
 G(t, \Gamma) = - \frac{1}{t} x^t_0 +O(x^{t+1}_0, \ln{x_0}),
\end{equation}
and for $t <1$ and $t$ is not an integer,
\begin{equation}
 G(t, \Gamma) = - \frac{1}{t} x^t_0 +O(x^{t+1}_0, 1).
\end{equation}
$G(t)$ is the well known Gamma function.

Next by using the above evaluations, we evaluate the integrals
defined by (\ref{doublegammadef}) which appear when we evaluate
the effects of the interactions between scalar fields on the 
final radiation energy density 
$\rho_{\alpha} = \sigma_{\alpha} / a^4$.
By expanding with respect to the small parameter $x_0$ by the partial
integration, we obtain the evaluations as follows.
We assume that $n_1 >0$.
For $n_2 >0$,
\begin{equation}
 G(n_1, n_2, \Gamma_1, \Gamma_2) = O(1),
\end{equation}
for $n_2 =0$,
\begin{equation}
 G(n_1, 0, \Gamma_1, \Gamma_2) = -
 \frac{1}{\Gamma^{n_1}_1} \ln{x_0} G(n_1)+O(1),
\end{equation}
and for $n_2 < 0$,
\begin{equation}
 G(n_1, n_2, \Gamma_1, \Gamma_2) = -
 \frac{1}{n_2}
 \frac{1}{\Gamma^{n_1}_1} x^{n_2}_0 G(n_1)
 +O(x^{n_1 + n_2}_0, x^{n_2 + 1}_0, \ln{x_0}).
\end{equation}

Finally we evaluate the incomplete Gamma function defined by
\begin{equation}
 G(t; x_1):= \int^{\infty}_{x_1} dx x^{t-1} e^{- x},
\end{equation}
for large $x_1$.
By partial integration, we obtain
\begin{equation}
 G(t; x_1) = x^{t-1}_1 e^{- x_1}+O(x^{t-2}_1 e^{- x_1}),
\end{equation}
for sufficiently large $x_1$.

\addtolength{\baselineskip}{-3mm}



\begin{thebibliography}{10}

\bibitem{Arnold}
{Arnold.V.I.},
Mathematical Methods of Classical Mechanics, (Springer, New York)
 (1978);
{ Arnold, V.I.} and {Avez, A.}
Probl\^^ {e}mes ergodiques de la m\'{e}canique classique, 
(Gauthier-Villars, Paris) (1967)

\bibitem{Bardeen1980}
{Bardeen, J.M.}, Phys. Rev. D {\bf 22}, 1882 (1980).

\bibitem{Bassett2000}
{Bassett, B.A.} and {Viniegra, F.}, Phys. Rev. D {\bf 62}, 043507
 (2000).

\bibitem{Dvali2004}
{Dvali, G.}, {Gruzinov, A.} and {Zaldarriga, M.}, Phys. Rev. D {\bf 69}, 023505
 (2004).

\bibitem{Finelli2000}
{Finelli, F.} and {Brandenberger, R.}, Phys. Rev. D{\bf 62}, 083502 (2000).

\bibitem{Gordon2001}
{ Gordon, C.}, {Wands, D.}, {Bassett, B.A.} and {Maartens, R.}, 
 Phys. Rev D{\bf 63} 123506 (2001).

\bibitem{Gupta2004}
{ Gupta, S.}, {Malik, K.A.} and {Wands, D.}, 
 Phys. Rev D{\bf 69} 063513 (2004).

\bibitem{Hamazaki1996}
{ Hamazaki, T.} and {Kodama, H.}, Prog. Theor. Phys. {\bf 96},1123--1146
  (1996).

\bibitem{Hamazaki2002}
{ Hamazaki, T.}, Phys. Rev. D {\bf 66}, 023529 (2002).

\bibitem{Hamazaki2004}
{ Hamazaki, T.}, Nucl. Phys. B {\bf 698},335--385 (2004).

\bibitem{Hosoya1984}
{ Hosoya, A.} and {Sakagami, M.}, 
 Phys. Rev D{\bf 29} 2228 (1984).

\bibitem{Kodama1984}
{Kodama, H. } and { Sasaki, M.}, Prog. Theor. Phys. Suppl. {\bf 78}, 1--166
  (1984).

\bibitem{Kodama1987}
{Kodama, H. } and { Sasaki, M.}, Int. J. Mod. Phys. {\bf A2}, 491
  (1987).

\bibitem{Kodama1996}
{Kodama, H. } and { Hamazaki, T.}, Prog. Theor. Phys. {\bf 96},949--970
  (1996).

\bibitem{Kodama1998}
{Kodama, H. } and { Hamazaki, T.}, Phys. Rev. D{\bf 57}, 7177--7185
  (1998).

\bibitem{Kofman1994}
 {Kofman, K.A.}, { Linde, A.D.} and {Starobinsky, A.A.}
Phys. Rev. Lett.{\bf 73}, 3195 (1994).

\bibitem{Kofman1997}
 {Kofman, K.A.}, { Linde, A.D.} and {Starobinsky, A.A.}
Phys. Rev. D{\bf 56}, 3258 (1997).

\bibitem{Komatsu2001}
 {Komatsu, E.} and {Spergel, D.N.}
Phys. Rev. D{\bf 63}, 063002 (2001).

\bibitem{Lyth2005}
 {Lyth, D.H.}, { Malik, K.A.} and {Sasaki, M.}
 JCAP. 0505, 004 (2005).

\bibitem{Malik2003}
{ Malik, K.A.}, {Wands, D.} and {Ungarelli, C.}, 
 Phys. Rev D{\bf 67} 063516 (2003).

\bibitem{Morikawa1986}
{ Morikawa, M.}, 
 Phys. Rev D{\bf 33} 3607 (1986).

\bibitem{Mukhanov1988}
{Mukhanov, V.F.}, Sov. phys.---JETP {\bf 67}, 1297--1302 (1988).

\bibitem{Mukhanov1992}
{Mukhanov, V.F.}, {Feldman, H.A.} and {Brandenberger, R.H.}
Phys. Rep. {\bf 215}, 203 (1992).

\bibitem{Nambu1997}
{Nambu, Y. } and { Taruya, A.}, Prog. Theor. Phys. {\bf 97},83--89
  (1997).

\bibitem{Nambu2006}
{Nambu, Y. } and { Araki, Y.}, Class. Quant. Grav. 23,511 (2006).

\bibitem{Polarski1992}
{ Polarski, D.} and {Starobinsky, A.A.}, Nucl. Phys. B {\bf 385}
 623 (1992).

\bibitem{Rigopoulos2003}
{Rigopoulos, G.I.} and {Shellard, E.P.S.}
Phys. Rev. D{\bf 68}, 123518 (2003).

\bibitem{Sasaki1986}
{Sasaki, M. }, Prog. Theor. Phys. {\bf 76},1036
  (1986). 

\bibitem{Sasaki1991}
{Sasaki, Y.} and {Yokoyama, J.}
Phys. Rev. D{\bf 44}, 970 (1991).

\bibitem{Sasaki1998}
{Sasaki, M. } and { Tanaka, T }, Prog. Theor. Phys. {\bf 99},763--782
  (1998). 

\bibitem{Shtanov1995}
{Shtanov, Y.}, {Traschen, J.} and {Brandenberger, R.H.}
Phys. Rev. D{\bf 51}, 5438 (1995).

\bibitem{Taruya1998}
{Taruya, A. } and { Nambu, Y.}, Phys. Lett. B428 37--43
  (1998). 

\bibitem{Traschen1990}
{Traschen, J.} and {Brandenberger, R.H.}
Phys. Rev. D{\bf 42}, 2491 (1990).


\bibitem{Wands2000}
{Wands, D.}, {Malik, K.A.}, {Lyth, D.H.} and {Liddle, A.R.}
Phys. Rev. D{\bf 62}, 043527 (2000).

\bibitem{Yokoyama2004}
{Yokoyama, J.}
Phys. Rev. D{\bf 70}, 103511 (2004).

\bibitem{Yoshida2004}
{ Yoshida, J.} and {Tsujikawa, S.}, 
Class. Quant. Grav. {\bf 23}, 353 (2006).

\bibitem{Zibin2001}
{Zibin, J.P.}, {Brandenberger, R} and {Scott, D}, 
 Phys. Rev. D{\bf 63}, 043511 (2001).

\end{thebibliography}
\end{document}